\documentclass[12pt]{article}

\usepackage{graphicx}

\def\gtwid{\mathrel{\raise.3ex\hbox{$>$\kern-.75em\lower1ex\hbox{$\sim$}}}}
\def\ltwid{\mathrel{\raise.3ex\hbox{$<$\kern-.75em\lower1ex\hbox{$\sim$}}}}
\def\square{\kern1pt\vbox{\hrule height 1.2pt\hbox{\vrule width 1.2pt\hskip 3pt
   \vbox{\vskip 6pt}\hskip 3pt\vrule width 0.6pt}\hrule height 0.6pt}\kern1pt}

\begin{document}

\begin{titlepage}

\begin{flushright}
UFIFT-QG-16-05
\end{flushright}

\vskip 2cm

\begin{center}
{\bf Determining Cosmology for a Nonlocal Realization of MOND}
\end{center}

\vskip 1cm

\begin{center}
M. Kim$^{1}$, M. H. Rahat$^{2}$, M. Sayeb$^{3}$, L. Tan$^{4}$, R. P. Woodard$^{5}$ and B. Xu$^{6}$
\end{center}

\vskip .5cm

\begin{center}
\it{Department of Physics, University of Florida,\\
Gainesville, FL 32611, UNITED STATES}
\end{center}

\vspace{1cm}

\begin{center}
ABSTRACT
\end{center}
We numerically determine the cosmological branch of the free function in a 
nonlocal metric-based modification of gravity which provides a relativistic
generalization of Milgrom's Modified Newtonian Dynamics. We are able to
reproduce the $\Lambda$CDM expansion history over virtually all of cosmic 
history, including the era of radiation domination during Big Bang 
Nucleosynthesis, the era of matter domination during Recombination, 
and most of the era of vacuum energy domination. The very late period
of $0 \leq z \ltwid 0.0880$, during which the model deviates from the
$\Lambda$CDM expansion history, is interesting because it causes the
current value of the Hubble parameter to be about 4.5\% larger than it 
would be for the $\Lambda$CDM model. This may resolve the tension between 
inferences of $H_0$ which are based on data from large redshift and 
inferences based on Hubble plots.

\begin{flushleft}
\end{flushleft}

\vskip .5cm

\noindent $^{1}$ pq8556@ufl.edu \hfill $^{4}$ billy@ufl.edu \\
\noindent $^{2}$ mrahat@ufl.edu \hfill $^{5}$ woodard@phys.ufl.edu \\
\noindent $^{3}$ sayebms1@ufl.edu \hfill $^{6}$ binxu@ufl.edu

\end{titlepage}

\section{Introduction}

There have been impressive confirmations of general relativity in the
solar system and from binary pulsars \cite{Will:2005va}. Most recently 
the existence of gravitational radiation has been established
\cite{Abbott:2016blz}. However, general relativity does not do such a
good job at explaining certain observed regularities of rotationally
supported galaxies:
\begin{itemize}
\item{{\it The Baryonic Tully-Fisher Relation} ($v^4_{\infty} = a_0
G M$) between the asymptotic rotational speed $v_{\infty}$ and the total
mass $M$ in baryons, where $a_0 \approx 1.2 \times 10^{-10}~{\rm m/s}^2$
\cite{McGaugh:2005qe};}
\item{{\it Milgrom's Law} that non-baryonic dark matter is required
when the gravitational acceleration from baryonic matter falls below $a_0$ \cite{Kaplinghat:2001me};}
\item{{\it Freeman's Law} ($\Sigma < \frac{a_0}{G}$) for the surface 
density $\Sigma$ \cite{Freeman:1970mx}; and}
\item{{\it Sancisi's Law} that features in the rotation curve are
correlated with features in the surface brightness \cite{Sancisi:2003xt}.}
\end{itemize}
Analogous regularities have been observed for pressure-supported galaxies
\cite{Sanders:2008iy}. These laws may signal undiscovered features in the 
way general relativity combines baryons with much more massive pools of 
dark matter to form cosmic structures. However, it is disconcerting that
increasingly sensitive experiments have so far failed to detect all this
dark matter \cite{Rosenberg:2015kxa,Akerib:2016lao,Akerib:2016vxi}.

In the absence of laboratory detection of dark matter it is worth while
examining the possibility that gravity is modified instead. Milgrom has 
proposed a particularly promising modification for what would be the static, 
weak field limit of such a theory \cite{Milgrom:1983ca,Milgrom:1983pn,
Milgrom:1983zz}. There is no question that Milgrom's MOdified Newtonian 
Dynamics (MOND) explains the various observed regularities of galactic
structure \cite{Sanders:2002pf,Famaey:2011kh}. The challenge is to extend
MOND to a fully relativistic theory that can be used to study the same
range of phenomena as general relativity. Two approaches have been followed:
\begin{itemize}
\item{Models in which other fields carry the MOND force 
\cite{Bekenstein:2004ne,Moffat:2005si}; and}
\item{Models in which an algebraic function of a nonlocal invariant 
of the metric is added to the Lagrangian --- $R \longrightarrow
R + \frac{a_0^2}{c^4} f_y(Z[g])$ --- \cite{Deffayet:2011sk}.}
\end{itemize}
The purpose of this paper is to specify $f_y(Z)$ for the latter approach. 

The modified gravity equations have been derived for a general metric
$g_{\mu\nu}$ and a general function $f_y(Z)$ \cite{Deffayet:2014lba,
Woodard:2014wia}. A crucially important point is that whereas the 
nonlocal invariant $Z[g]$ is positive for gravitationally bound systems, 
it is typically negative for cosmological systems in which time dependence 
is more significant than spatial dependence. Explaining the regularities
of galaxies fixes the algebraic function for positive arguments 
\cite{Deffayet:2011sk}, 
\begin{equation}
f_y(Z) = \frac12 Z - \frac16 Z^{\frac32} + O(Z^2) \; .
\end{equation}
Preserving solar system tests (and those at strong fields) requires that
the function fall off for large, positive $Z$. For example, the following
simple form would suffice \cite{Deffayet:2011sk},
\begin{equation}
Z > 0 \qquad \Longrightarrow \qquad f_y(Z) = \frac12 Z \exp\Bigl[ -\frac13
\sqrt{Z} \, \Bigr] \; .
\end{equation}
But how the function depends on negative $Z$ must be regarded as a free 
parameter at this stage. 

The purpose of this paper is to show that the function $f_y(Z)$ can be
chosen so as to reproduce the $\Lambda$CDM expansion history --- which
is itself just a model, with free parameters --- over almost the full 
range of cosmic evolution. It turns out that the nonlocal invariant $Z$ 
fails to be monotonic at very late times, which means that our nonlocal 
realization of MOND inevitably deviates from $\Lambda$CDM cosmology. 
However, this deviation is serendipitous because it makes the current 
value of the Hubble parameter about 4.5\% larger, which has the potential 
to explain why high redishift determinations of $H_0$ give a smaller value 
than low redshift determinations.

In section 2 we review the model, defining the nonlocal invariant $Z[g]$
and specializing the field equations to a general, spatially flat cosmology. 
Section 3 assumes the $\Lambda$CDM expansion history and uses this to derive
the form of $f_y(Z)$ must take for large negative $Z$ and for small negative 
$Z$. Section 4 presents a numerical solution of the full problem, which 
includes the failure of $Z$ to be monotonic at very late times and the 
resulting deviation from $\Lambda$CDM cosmology. What all this means is
discussed in section 5.

\section{The Model}

The purpose of this section is to briefly review the nonlocal metric
realization of MOND \cite{Deffayet:2011sk,Deffayet:2014lba}. We begin by 
defining the full invariant $Z[g](x)$, then give the general field equations. 
The section closes by specializing $Z[g](x)$ and the field equations to the
homogeneous and isotropic geometry of cosmology. 

The gravitational Lagrangian takes the form,
\begin{equation}
\mathcal{L} = \frac{c^4}{16 \pi G} \Biggl\{ R + \frac{a^2_0}{c^4} 
f_y\Bigl( Z[g]\Bigr) \Biggr\} \sqrt{-g} \; . \label{LMOND}
\end{equation}
The dimensionless nonlocal invariant $Z[g]$,
\begin{equation}
Z[g] \equiv \frac{4 c^4}{a_0^2} \, g^{\mu\nu} \Bigl[ \partial_{\mu} 
\frac1{\square} R_{\alpha\beta} u^{\alpha} u^{\beta} \Bigr] 
\Bigl[ \partial_{\nu} \frac1{\square} R_{\rho\sigma} u^{\rho} 
u^{\sigma} \Bigr] \; , \label{Zdef}
\end{equation}
is constructed using three geometrical quantities:
\begin{enumerate}
\item{The inverse scalar d'Alembertian,
\begin{equation}
\square \equiv \frac1{\sqrt{-g}} \, \partial_{\mu} \Bigl( \sqrt{-g} 
\, g^{\mu\nu} \partial_{\nu} \Bigr) \; , \label{box}
\end{equation}
where the inverse is defined using retarded boundary conditions.\footnote{
It might be preferable to define the initial derivative of $\Phi = 
\frac1{\square}$ of anything as $-g^{\mu\nu} \partial_{\mu} \Phi 
\partial_{\nu} \Phi = {\rm Constant}$ \cite{Tsamis:2014kda}. For late 
times this improvement is not needed.}}
\item{A normalized timelike 4-velocity $u^{\mu}[g](x)$,
\begin{equation}
\chi[g](x) \equiv -\frac1{\square} \, 1 \qquad \Longrightarrow \qquad 
u^{\mu}[g] \equiv \frac{-g^{\mu\nu} \partial_{\nu} \chi[g]}{\sqrt{-
g^{\alpha\beta} \partial_{\alpha} \chi[g] \partial_{\beta} \chi[g]}} \; .
\label{umu}
\end{equation}}
\item{The Ricci tensor $R_{\mu\nu}$.}
\end{enumerate}

At this point a digression is in order to comment on the probable genesis of 
nonlocality. We believe that fundamental theory is local, but the nonlocal 
Lagrangian (\ref{LMOND}) represents the gravitational vacuum polarization of 
the vast ensemble of infrared gravitons created by primordial inflation. 
These gravitons were not present at the beginning of inflation, and their 
wave lengths do not extend to arbitrarily small scales. This in no way 
changes the purely phenomenological status of the nonlocal model (\ref{LMOND}), 
but it does explain two of the model's features which would otherwise seem 
absurd \cite{Woodard:2014wia}:
\begin{itemize}
\item{That there is an initial time at which to specify the initial conditions
of the inverse d'Alembertian; and}
\item{That MOND corrections affect large scales but not small scales.}
\end{itemize}

The simplest way to present the general field equations is by introducing 
nondynamical, auxiliary scalars after the method of Nojiri and Odintsov
\cite{Nojiri:2007uq}. Localizing the Lagrangian (\ref{LMOND}) requires
four such scalars \cite{Deffayet:2014lba},
\begin{eqnarray}
\lefteqn{\mathcal{L} = \frac{c^4}{16\pi G} \Biggl\{ R + \frac{a_0^2}{c^4} 
f_y\Bigl(\frac{g^{\mu\nu} \partial_{\mu} \phi \partial_{\nu} \phi}{
c^{-4} a_0^2} \Bigr) } \nonumber \\
& & \hspace{2cm} - \Bigl[ \partial_{\mu} \xi \partial_{\nu} \phi
g^{\mu\nu} \!+\! 2 \xi R_{\mu\nu} u^{\mu} u^{\nu} \Bigr] - \Bigl[
\partial_{\mu} \psi \partial_{\nu} \chi g^{\mu\nu} \!-\! \psi\Bigr]
\Biggr\} \sqrt{-g} \; , \qquad \label{localMOND}
\end{eqnarray}
where $D_{\mu}$ denotes the covariant derivative operator and $u^{\mu}[g]$
is defined in terms of the scalar $\chi$ according to the right hand side
of (\ref{umu}). The auxiliary scalars must not be considered as independent 
degrees of freedom because two of them would be ghosts \cite{Deser:2013uya,
Woodard:2014iga}. The four scalars are rather nonlocal functionals of the 
metric defined by solving their equations of motion with retarded boundary 
conditions ,
\begin{eqnarray}
\phi[g] = \frac2{\square} R_{\alpha\beta} u^{\alpha} u^{\beta}  
&\!\! , \!\!& \chi[g] = -\frac1{\square} 1 \; , \label{scalars1} \\
\xi[g] = \frac2{\square} D^{\mu} \Biggl[ \partial_{\mu} \phi f'_y\Bigl(
\frac{ g^{\rho\sigma} \partial_{\rho} \phi \partial_{\sigma} \phi}{c^{-4} a_0^2}
\Bigr) \Biggr] &\!\! , \!\! & \psi[g] = \frac{4}{\square} D_{\mu} 
\Biggl[ \frac{ \xi (g^{\mu\rho} \!+\! u^{\mu} u^{\rho}) u^{\sigma} 
R_{\rho\sigma}}{\sqrt{-g^{\alpha\beta} \partial_{\alpha} \chi \partial_{\beta}
\chi}}\Biggr] . \quad \label{scalars2}
\end{eqnarray} 
The modified gravitational field equations are \cite{Deffayet:2014lba},
\begin{eqnarray}
\lefteqn{ R_{\mu\nu} + \frac12 g_{\mu\nu} \Bigl[ -R -\frac{a_0^2}{c^4}
f_y + g^{\rho\sigma} \Bigl( \partial_{\rho} \xi \partial_{\sigma} \phi 
\!+\! \partial_{\rho} \psi \partial_{\sigma} \chi\Bigr) + 2 \xi 
u^{\rho} u^{\sigma} R_{\rho\sigma} - \psi\Bigr] } \nonumber \\
& & \hspace{-.3cm} + \partial_{\mu} \phi \partial_{\nu} \phi f_y' -
\partial_{(\mu} \xi \partial_{\nu)} \phi - \partial_{(\mu} \psi
\partial_{\nu)} \chi - 2 \xi \Bigl[ 2 u_{(\mu} u^{\alpha} R_{\nu ) \alpha}
\!+\! u_{\mu} u_{\nu} u^{\alpha} u^{\beta} R_{\alpha\beta} \Bigr]
\nonumber \\
& & \hspace{.3cm} - \Bigl[ \square (\xi u_{\mu} u_{\nu}) +
g_{\mu\nu} D_{\alpha} D_{\beta} (\xi u^{\alpha} u^{\beta}) -
2 D_{\alpha} D_{(\mu} (\xi u_{\nu)} u^{\alpha} ) \Bigr] 
= \frac{8\pi G}{c^4} \, T_{\mu\nu} \; . \qquad \label{fulleqns}
\end{eqnarray}

It remains to specialize relations (\ref{scalars1}-\ref{fulleqns}) to the
homogeneous, isotropic and spatially flat geometry relevant to cosmology,
\begin{equation}
g_{\mu\nu} dx^{\mu} dx^{\nu} = -c^2 dt^2 + a^2(t) d\vec{x} \cdot
d\vec{x} \quad \Longrightarrow \quad H(t) \equiv \frac{\dot{a}}{a} 
\quad , \quad \epsilon(t) \equiv -\frac{\dot{H}}{H^2} \; . \label{FLRW}
\end{equation}  
In this geometry the auxiliary scalars become \cite{Deffayet:2014lba},
\begin{eqnarray}
\phi(t) & = & -6 \!\! \int_{t_i}^t \!\! \frac{dt'}{a^3(t')} \!\!
\int_{t_i}^{t'} \!\! dt'' a^3(t'') H^2(t'') \Bigl[\epsilon(t'') \!-\!
1 \Bigr] \; \Longrightarrow \; Z(t) = -\frac{\dot{\phi}^2(t)}{
c^{-2} a_0^2} \; , \qquad \label{FLRWphi} \\
\chi(t) & = & \!\! \int_{t_i}^t \!\! \frac{dt'}{a^3(t')} \!\!
\int_{t_i}^{t'} \!\! dt'' a^3(t'') \qquad \Longrightarrow \qquad
u^{\mu}(t) = \delta^{\mu}_0 \; , \qquad \label{FLRWchi} \\
\xi(t) & = & 2 \!\! \int_{t_i}^t \!\! dt' \dot{\phi}(t') f_y'\Bigl(
Z(t') \Bigr) \qquad , \qquad \psi(t) = 0 \; , \label{FLRWxipsi}
\end{eqnarray}
where $t_i$ is the initial time. The gravitational field equations are 
\cite{Deffayet:2014lba},
\begin{eqnarray}
3 H^2 \!+\! \frac{a_0^2}{2 c^2} f_y(Z) \!+\! 3 H \dot{\xi} \!+\! 6 H^2 \xi 
& = & \frac{8\pi G}{c^2} \, \rho \; , \label{Fried1} \\
-2\dot{H} \!-\! 3H^2 \!-\! \frac{a_0^2}{2 c^2} \, f_y(Z) \!-\! \ddot{\xi} 
\!-\! \Bigl(\frac{\dot{\phi}}2 \!+\! 4 H\Bigr) \dot{\xi} \!-\! 
\Bigl(4\dot{H} \!+\! 6 H^2\Bigr) \xi  & = &  \frac{8\pi G}{c^2} \, p
\; , \qquad \label{Fried2}
\end{eqnarray}
where $\rho(t)$ is the energy density and $p(t)$ is the pressure.

\section{Asymptotic Analysis}

The aim of this paper is to choose the function $f_y(Z)$ so that equation 
(\ref{Fried1}) reproduces the $\Lambda$CDM expansion history without including 
dark matter in the energy density $\rho(t)$. The purpose of this particular 
section is to accomplish that task analytically for large negative $Z$ and for
small negative $Z$. We begin by defining the $\Lambda$CDM model, with the usual
choice of redshift $z$ as the time parameter. We then present exact equations
for the nonlocal invariant $Z$ and the algebraic function $f_y(Z)$. Solving 
these exact equations for large and small negative $Z$ completes the asymptotic
analysis of the section. 

\subsection{The $\Lambda$CDM Model}

Cosmologists employ the cosmological redshift $z$ as the time variable,
\begin{equation}
1 + z \equiv \frac{a_0}{a(t)} \qquad \Longrightarrow \qquad 
\frac{dz}{1 \!+\! z} = -H dt \; , \label{redshift}
\end{equation}
where $a_0 \equiv a(t_0)$ is the value of the scale factor at the current 
time $t_0$. In the interest of simplicity we will abuse the notation slightly 
by referring to standard quantities $H(z)$, $\epsilon(z)$ and $Z(z)$ as
functions of redshift. 

The $\Lambda$CDM model is defined by assuming the Hubble parameter is,
\begin{equation}
H(z) \equiv H_0 \sqrt{\Omega_{\rm rad} (1 \!+\! z)^4 + \Omega_{\rm mat} 
(1 \!+\! z)^3 + \Omega_{\Lambda}} \equiv H_0 \!\times\! \widetilde{H}(z) 
\; , \label{Htilde}
\end{equation}
where the parameters $\Omega_{\rm r}$, $\Omega_{\rm m}$ and 
$\Omega_{\Lambda}$ are \cite{Ade:2015xua},
\begin{equation}
\Omega_{\rm r} = 0.0000916 \quad ,\quad \Omega_{\rm m} = 0.309 \quad , \quad
\Omega_{\Lambda} = 0.691 \; . \label{Omegas}
\end{equation}
The first slow roll parameter of the $\Lambda$CDM model is,
\begin{equation}
\epsilon(z) \equiv \frac{2 \Omega_{\rm r} (1 \!+\! z)^4 +
\frac32 \Omega_{\rm m} (1 \!+\! z)^3}{ \Omega_{\rm r}
(1 \!+\! z)^4 + \Omega_{\rm m} (1 \!+\! z)^3 + \Omega_{\Lambda}}
\; . \label{epsilon}
\end{equation}

Without dark matter the energy density of equation (\ref{Fried1}) is,
\begin{equation}
\frac{8 \pi G}{c^2} \, \rho = 3 H_0^2 \Bigl[ \Omega_{\rm r} (1 \!+\! z)^4
+ \Omega_{\rm b} (1 \!+\! z)^3 + \Omega_{\Lambda}\Bigr] \; . \label{LCDMrho}  
\end{equation}
The baryonic fraction of the critical density is \cite{Ade:2015xua},
\begin{equation}
\Omega_{\rm b} = 0.0486 \; . \label{baryons}
\end{equation}
The missing mass is \cite{Ade:2015xua},
\begin{equation}
\frac{8 \pi G}{c^2} \, \rho - 3 H^2 = -3 H_0^2 \!\times\! \Omega_{\rm c} 
(1 \!+\! z)^3 \qquad , \qquad \Omega_{\rm c} = 0.259 \; .
\end{equation}

\subsection{Exact Relations for $Z(z)$ and $f_y(Z)$.}

MOND corrections depend on the dimensionless ratio of $c$ and $H_0$ to 
$a_0$,
\begin{equation}
\alpha \equiv \frac{6 c H_0}{a_0} \simeq 33 \; . \label{alpha}
\end{equation}
Because the initial redshift is effectively infinite the function $Z(z)$ 
is,
\begin{equation}
\sqrt{-Z(z)} = \alpha (1 \!+\! z)^3 \!\! \int_{z}^{\infty} \!\!
\frac{dz'}{(1 \!+\! z')^4} \, \frac{\Omega_{\rm r} (1 \!+\! z')^4
+ \frac12 \Omega_{\rm m} (1 \!+\! z')^3 - \Omega_{\Lambda}}{\sqrt{
\Omega_{\rm r} (1 \!+\! z')^4 + \Omega_{\rm m} (1 \!+\! z')^3 +
\Omega_{\Lambda}}} \; . \label{ztoZ}
\end{equation}
The right hand side of expression (\ref{ztoZ}) is obviously positive 
and monotonically increasing for large $z$. For small $z$ the 
$-\Omega_{\Lambda}$ term in the numerator of the integrand changes 
the sign of the integrand, and we will presently see that this causes
the integral to vanish at $z_{*} \simeq 0.0880$. Because $f_y$ is a
function of $Z$, rather than $z$, we cannot enforce the $\Lambda$CDM
expansion history for the small region $0 < z < z_{*}$, so we 
concentrate on the region $z_{*} < z < \infty$ over which the 
transformation from $z$ to $Z(z)$ is one-to-one. In this region the
modified Friedman equation (\ref{Fried1}) can be written,
\begin{equation}
\frac{f_y(Z)}{2 \alpha^2 \widetilde{H}^2}  - \frac{ \sqrt{-Z} \, 
f'_y(Z)}{\alpha \widetilde{H}} + \int_{Z}^{\infty} \frac{f'_y(Z') dZ'
}{(1 \!+\! z') \frac{d\sqrt{-Z'}}{d z'} \times\! \alpha \widetilde{H}} 
= -\frac{\Omega_c}{12} \frac{(1 \!+\! z)^3}{\widetilde{H}^2} \; . 
\label{exacteqn}
\end{equation}

\subsection{Results for large negative $Z$}

The integration in expression (\ref{ztoZ}) can be reduced to elliptic 
integrals but the result is not especially useful. However, large
negative $Z$ corresponds to large $z$, in which case one can neglect
$\Omega_{\Lambda}$ to obtain a useful expression,
\begin{eqnarray}
\sqrt{-Z} & \simeq & \alpha (1 \!+\! z)^3 \!\! \int_{z}^{\infty} \!\!
\frac{dz'}{(1 \!+\! z')^4} \, \frac{\Omega_{\rm r} (1 \!+\! z')^4
+ \frac12 \Omega_{\rm m} (1 \!+\! z')^3}{\sqrt{\Omega_{\rm r} 
(1 \!+\! z')^4 + \Omega_{\rm m} (1 \!+\! z)^3}} \; , \\
& = & \frac{\alpha \sqrt{\Omega_{\rm r}} \, (1 \!+\! z)^3}{1 \!+\! 
z_{\rm eq}} \Biggl\{ \frac13 \Bigl[ 1 \!+\! \frac{1 \!+\! z_{\rm eq}}{
1 \!+\! z}\Bigr]^{\frac32} + \Bigl[1 + \frac{1 \!+\! z_{\rm eq}}{1 
\!+\! z}\Bigr]^{\frac12} - \frac43\Biggr\} , \label{Zall} \\
& = & \alpha \sqrt{\Omega_{\rm r}} \, (1 \!+\! z)^2 \Biggl\{1 +
\sum_{n=2}^{\infty} \frac{(n \!-\! 1) (2n \!-\! 3)!!}{(n \!+\! 1)!}
\Bigl[- \frac12 \frac{1 \!+\! z_{\rm eq}}{1 \!+\! z}\Bigr]^{n} \Biggr\} 
, \qquad \label{largeZ}
\end{eqnarray}
where $z_{\rm eq} \equiv \frac{\Omega_{\rm m}}{\Omega_{\rm r}} \simeq
3370$ is the redshift of matter-radiation equality. 

Although the series (\ref{largeZ}) converges there is no point to 
proceeding higher than $n=3$ because the $n=4$ term has the same 
strength as the neglected $\Omega_{\Lambda}$ contributions. The various 
expansions look simpler in terms of the quantities,
\begin{equation}
\beta \equiv \sqrt{\alpha} \, (1 \!+\! z_{\rm eq}) \qquad , \qquad 
\zeta \equiv -\frac{Z}{\Omega_{\rm r}} \qquad \Longrightarrow \qquad 
\frac{d}{d Z} = -\frac1{\Omega_{\rm r}} \frac{d}{d\zeta} \; .
\end{equation}
From (\ref{largeZ}) we find,
\begin{eqnarray}
\widetilde{H} & = & \frac{\sqrt{\Omega_{\rm r}}}{\alpha} \, 
\zeta^{\frac12} \Biggl\{ 1 + \frac12 \frac{\beta}{\zeta^{\frac14}} 
- \frac16 \frac{\beta^2}{\zeta^{\frac12}} + \frac{3}{32} 
\frac{\beta^3}{\zeta^{\frac34}} + O\Bigl( \frac{\beta^4}{\zeta}\Bigr) 
\Biggr\} , \label{largeZH} \\
(1 \!+\! z) \frac{d \sqrt{-Z}}{dz} & = & 2 \sqrt{\Omega_{\rm r}} \,
\zeta^{\frac12} \Biggl\{1 - \frac1{24} \frac{\beta^2}{\zeta^{\frac12}}
+ \frac{3}{64} \frac{\beta^3}{\zeta^{\frac34}} + O\Bigl( 
\frac{\beta^4}{\zeta}\Bigr) \Biggr\} , \label{largeZdZdz} \\
\frac{(1 \!+\! z)^3}{\widetilde{H}^2} & = & \frac{\sqrt{\alpha}}{
\Omega_{\rm r}} \, \zeta^{-\frac14} \Biggl\{ 1 - \frac{\beta}{
\zeta^{\frac14}} + \frac{49}{48} \frac{\beta^2}{\zeta^{\frac12}} -
\frac{203}{192} \frac{\beta^3}{\zeta^{\frac34}} + O\Bigl( 
\frac{\beta^4}{\zeta} \Bigr) \Biggr\} . \label{largeZsource}
\end{eqnarray}
Substituting these relations in (\ref{exacteqn}) implies,
\begin{equation}
f_y(Z) = -\frac{\sqrt{\alpha} \, \Omega_{\rm c}}{33} \, \zeta^{\frac34}
\Biggl\{ 1 - \frac{\beta}{\zeta^{\frac14}} + \frac{155}{176} 
\frac{\beta^2}{\zeta^{\frac12}} - \frac{625}{768} \frac{\beta^3}{
\zeta^{\frac34}} + O\Bigl( \frac{\beta^4}{\zeta} \Bigr) \Biggr\} . 
\label{largeZansatz}
\end{equation}

Of course relation (\ref{exacteqn}) is an inhomogeneous, first order
integro-differential equation so its solution is ambiguous up to the
addition of a homogeneous solution. For large $Z$ this solution takes
the form of a constant times,
\begin{equation}
f_{\rm h}(Z) = \zeta^{\frac12 - \frac{\sqrt{3}}{2}} \Biggl\{\!1 - 
\frac25 \Bigl(1 \!+ \! \sqrt{3} \Bigr) \frac{\beta}{\zeta^{\frac14}} \!+
\Bigl[ \frac{67}{90} \!-\! \frac{7}{24} \Bigl(1 \!-\! \sqrt{3}\Bigr)
\Bigr] \frac{\beta^2}{\zeta^{\frac12}} \!+ O\Bigl( \frac{\beta^3}{
\zeta^{\frac34}} \Bigr) \! \Biggr\} . \label{homogeneous}
\end{equation}
The coefficient of $f_{\rm h}(Z)$ is not fixed by the asymptotic 
expansion (\ref{largeZansatz}) so we can use it to impose the condition 
that $f_y(Z)$ vanishes at $Z=0$.

\subsection{Results for small negative $Z$}

As $z$ is reduced it eventually becomes invalid to ignore 
$\Omega_{\Lambda}$. A reasonable measure of this point is when
the energy density in radiation becomes equal to the vacuum energy
density,
\begin{equation}
\Omega_{\rm r} (1 \!+\! z_{\rm tr})^4 = \Omega_{\Lambda} \qquad 
\Longrightarrow \qquad z_{\rm tr} = \Bigl( \frac{\Omega_{\Lambda}}{
\Omega_{\rm r}} \Bigr)^{\frac14} - 1 \simeq 8.32 \; . \label{ztrans}
\end{equation}
At this point the energy density in matter is still greatly predominant,
\begin{equation}
\Omega_{\rm r} (1 \!+\! z_{\rm tr})^4 = \Omega_{\Lambda} = 0.691 \ll
\Omega_{\rm m} (1 \!+\! z_{\rm tr})^3 \simeq 250 \; .
\end{equation}
For $z < z_{\rm tr}$ we therefore make only a small error by writing,
\begin{equation}
\sqrt{-Z} \simeq \alpha (1 \!+\! z)^3 \Biggl\{ 
\int_{z_{\rm tr}}^{\infty} \!\! \frac{dz'}{(1 \!+\! z')^2} 
\frac{ \Omega_{\rm r} \!+\! \frac{\Omega_{\rm m}}{2 (1 \!+\! z')}}{
\sqrt{\Omega_{\rm r} \!+\! \frac{\Omega_{\rm m}}{1 \!+\! z'}}} + 
\int_{z}^{z_{\rm tr}} \!\! \frac{dz'}{(1 \!+\! z')^{\frac52}} 
\frac{\frac12 \Omega_{\rm m} \!-\! \frac{ \Omega_{\Lambda}}{(1 \!+\! 
z')^3}}{\sqrt{ \Omega_{\rm m} \!+\! \frac{\Omega_{\Lambda}}{(1 \!+\! 
z')^3}}} \Biggr\} . \label{smallZapprox}
\end{equation}
The two integrals of (\ref{smallZapprox}) can be expressed in terms of,
\begin{equation}
Y \equiv \sqrt{ \frac{1 \!+\! z_{\rm tr}}{1 \!+\! z_{\rm eq}}} \simeq
0.0526 \; , \; 1 + z_{\Lambda} \equiv \Bigl( \frac{\Omega_{\Lambda}}{
\Omega_{\rm m}}\Bigr)^{\frac13} \simeq 1.31 \; , \; y(z) \equiv \Bigl( 
\frac{1 \!+\! z_{\Lambda}}{1 \!+\! z} \Bigr)^{\frac32} \; .
\end{equation}
The result for (\ref{smallZapprox}) is,
\begin{eqnarray}
\lefteqn{\sqrt{-Z} \simeq \frac{\alpha \sqrt{\Omega_{\Lambda}}}{3 y^2} 
\Biggl\{ (2 Y \!+\! 5 Y^3) \sqrt{1 \!+\! Y^2} - 4 Y^4 } \nonumber \\
& & \hspace{2.5cm} - 2 \ln\Bigl[Y \!+\! \sqrt{1 \!+\! Y^2} \, \Bigr] 
- y \sqrt{1 \!+\! y^2} + 
2 \ln\Bigl[y \!+\! \sqrt{1 \!+\! y^2} \, \Bigr] \Biggr\} , \quad 
\label{smallform} \\
& & \hspace{-.5cm} = \frac{\alpha \sqrt{\Omega_{\Lambda}}}{3 y^2} 
\Biggl\{\! y_* \sqrt{1 \!+\! y_*^2} \!-\! 2 \ln\Bigl[y_* \!\!+\!\! 
\sqrt{1 \!+\! y_*^2} \, \Bigr] \!-\! y \sqrt{1 \!+\! y^2} \!+\! 
2 \ln\Bigl[y \!+\!\! \sqrt{1 \!+\! y^2} \, \Bigr] \!\Biggr\} , \quad
\label{smallZform}
\end{eqnarray} 
where $y_* \equiv y(z_*) \simeq 1.318$ and $z_* \simeq 0.0880$ is the 
redshift at which $Z$ vanishes.

It is best to expand (\ref{smallZform}), and everything else, in powers 
of the parameter $\Delta y \equiv y - y_{*} < 0$,
\begin{equation}
\sqrt{-Z} = \frac{\alpha \sqrt{\Omega_{\Lambda}}}{3 y_{*}^2} \Biggl\{ 
\Bigl( \frac{2 y_{*}^2 \!-\! 1}{\sqrt{1 \!+\! y_{*}^2}} \Bigr) (-\Delta y)
- \frac{(4 \!+\! y_{*}^2 \!-\! 6 y_{*}^4)}{(1 \!+\! y_{*}^2)^{\frac32}}
\frac{(-\Delta y)^2}{2 y_{*}} + O\Bigl( \Delta y^3\Bigr) \Biggr\} . 
\label{Ztoy}
\end{equation}
Inverting relation (\ref{Ztoy}) gives,
\begin{equation}
-\Delta y = \frac{3 y_{*}^2 \sqrt{1 \!+\! y_{*}^2}}{2 y_{*}^2 \!-\! 1} 
\sqrt{\frac{-Z}{\alpha^2 \Omega_{\Lambda}}} \Biggl\{ 1 + \frac{3 y_{*}
(4 \!+\! y_{*}^2 \!-\! 6 y_{*}^4)}{2 \sqrt{1 \!+\! y_{*}^2} \, (2 y_{*}^2
\!-\! 1)^2} \sqrt{\frac{-Z}{\alpha^2 \Omega_{\Lambda}}} + O(Z) \Biggr\} .
\label{ytoZ}
\end{equation}
From expression {(\ref{ytoZ}) we see that the various expansions will be
simpler when expressed in terms of the variable $\mathcal{Z}$,
\begin{equation}
\mathcal{Z} \equiv -\frac{Z}{\alpha^2 \Omega_{\Lambda}} \qquad 
\Longrightarrow \qquad \frac{d}{dZ} = -\frac1{\alpha^2 \Omega_{\Lambda}}
\frac{d}{d \mathcal{Z}} \; .
\end{equation} 

Before expanding in $\mathcal{Z}$ we will give the relevant expressions 
(exact up to ignoring $\Omega_{\rm r}$) in terms of $y$,
\begin{eqnarray}
\widetilde{H} = \sqrt{\Omega_{\Lambda}} \sqrt{1 \!+\! \frac1{y^2}} 
\quad & , & \quad \frac{(1 \!+\! z)^3}{\widetilde{H}^2} = 
\frac1{\Omega_{\rm m}} \, \frac1{1 \!+\! y^2} \; , 
\label{smallZHsource} \\
(1 \!+\! z) \frac{d \sqrt{-Z}}{dz} & = & \alpha \sqrt{\Omega_{\Lambda}}
\Biggr[ \frac{1 \!-\! \frac1{2 y^2}}{\sqrt{1 \!+\! \frac1{y^2}}} + 3
\sqrt{\mathcal{Z}} \Biggr] . \label{smallZderiv}
\end{eqnarray}
Substituting relations (\ref{smallZHsource}-\ref{smallZderiv}) into 
(\ref{exacteqn}) implies,
\begin{equation}
\frac{\frac12 f_y}{1 \!+\! \frac1{y^2}} + \frac{ \sqrt{\mathcal{Z}} \, 
\frac{d f_y}{d \mathcal{Z}}}{\sqrt{1 \!+\! \frac1{y^2}}} + 
\int_{\mathcal{Z}}^{\infty} \frac{ \frac{d f_y}{d\mathcal{Z}'} 
d\mathcal{Z}'}{1 \!-\! \frac1{2 {y'}^2} \!+\! 3 \sqrt{1 \!+\! 
\frac1{{y'}^2}} \sqrt{\mathcal{Z}'}} = 
-\frac{\alpha^2 \Omega_{\rm c}}{12} \frac{(1 \!+\! z_{\Lambda})^3}{1 
\!+\! y^2} \; . \label{yeqn}
\end{equation}
The integral in (\ref{yeqn}) can be expressed in terms of a constant
minus a part which vanishes with $\mathcal{Z}$,
\begin{equation}
\int_{\mathcal{Z}}^{\infty} \!\! d\mathcal{Z}' = \int_{0}^{\infty} \!\!
d\mathcal{Z}' - \int_{0}^{\mathcal{Z}} \!\! d\mathcal{Z}' \equiv K
- \int_{0}^{\mathcal{Z}} \!\! d\mathcal{Z}' \; .
\end{equation}
The constant $K$ is determined by the requirement that $f_y(Z)$ vanishes
at $Z = 0$. There is no simple expression for it but we determined 
(numerically) that its value is $K \simeq -35.2$.

Equation (\ref{yeqn}) suggests that the small $Z$ form of $f_y(Z)$ is,
\begin{equation}
f_y(Z) = -\frac{\alpha^2 \Omega_{\rm c}}{12} \!\times\! (1 \!+\! z_{\Lambda})^3
\! \times \! \sqrt{\mathcal{Z}} \, \Bigl[ A + B \sqrt{\mathcal{Z}} + 
O( \mathcal{Z}) \Bigr] \; . \label{smallZansatz}
\end{equation} 
Using this in (\ref{yeqn}), with $k \equiv -\frac{12 K}{\alpha^2 \Omega_{c} 
(1 + z_{\Lambda})^3} \simeq 0.670$, produces the equation,
\begin{eqnarray}
\lefteqn{\frac{ [\frac12 A \sqrt{\mathcal{Z}} \!+\! \frac12 B \mathcal{Z} 
\!+\! \dots]}{1 \!+\! \frac1{y^2}} + \frac{[\frac12 A \!+\! B 
\sqrt{\mathcal{Z}} \!+\! \dots]}{\sqrt{1 \!+\! \frac1{y^2}}} + k - 
\frac1{1 \!+\! y^2} } \nonumber \\
& & \hspace{6cm} = \int_{0}^{\mathcal{Z}} \!\! \frac{ 
[\frac{A}{2 \sqrt{\mathcal{Z}'}} \!+\! B \!+\! \dots] d\mathcal{Z}'}{1 
\!-\! \frac1{2 {y'}^2} \!+\! 3 \sqrt{1 \!+\! \frac1{{y'}^2}} \, 
\sqrt{\mathcal{Z}'} } \; . \qquad \label{newy}
\end{eqnarray}
The coefficient $A$ is determined by the order $\mathcal{Z}^0$ term in
(\ref{newy}),
\begin{equation}
A = \frac{2 [1 \!-\! (1 \!+\! y_{*}^2) k]}{y_{*} \sqrt{1 \!+\! y_{*}^2}}
\simeq -0.764 \; . 
\end{equation}
The coefficient $B$ comes from the order $\sqrt{\mathcal{Z}}$ term in 
(\ref{newy}),
\begin{equation}
B = \frac{8 \!-\! 2 (4 \!+\! y_{*}^2) k}{(2 y_{*}^2 \!-\! 1)} \simeq
+0.127 \; .
\end{equation}

\section{Numerical Analysis}

The purpose of this section is to numerically determine the function
$f_y(Z)$ for negative $Z$, checking each step against the analytic results
of the previous section. Because $Z(z)$ vanishes for $z = z_{*} \simeq
0.0880$ one cannot exactly reproduce the $\Lambda$CDM model in the range 
$0 \leq z < z_{*}$ so we must also quantify the extent that the modified
Friedman equation (\ref{Fried1}) causes the expansion history to deviate
from the $\Lambda$CDM expansion history. Our technique for finding $f_y(Z)$
is to first determine it and $Z(z)$ numerically as functions of $z$ for
$z \geq z_{*}$. We then invert the relation between $z$ and $Z$ to construct
$z(Z)$, and use this to find $f_y$ as a function of $Z$, both numerically 
and by fitting to an analytic function. The section closes with a numerical 
evolution of the modified Friedman equation (\ref{Fried1}) in the range 
$0 < z < z_{*}$ to determine how much the Hubble parameter deviates from 
that of the $\Lambda$CDM model.    

\subsection{Converting to $z$ and factoring out $\alpha$ and $\Omega_{\rm c}$}

\begin{figure}[ht]
\includegraphics[width=6cm,height=4.8cm]{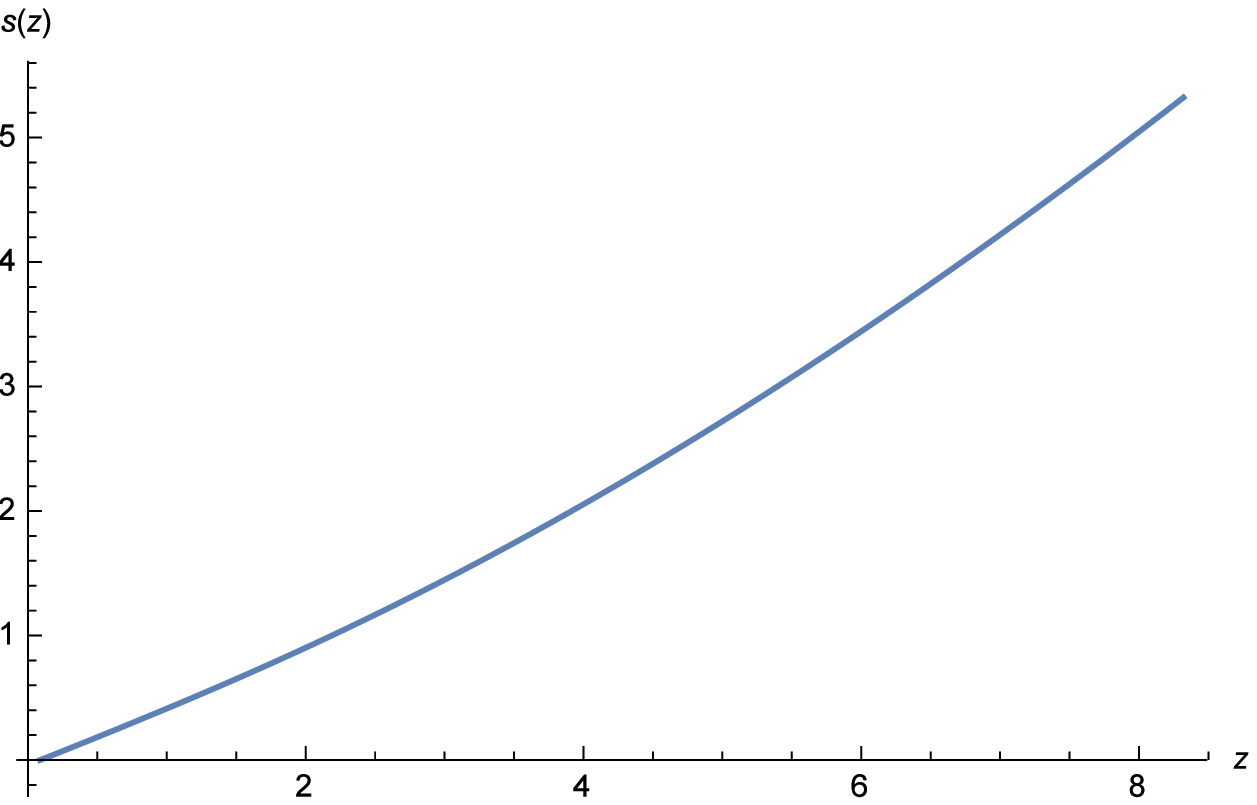}
\hspace{1cm}
\includegraphics[width=6cm,height=4.8cm]{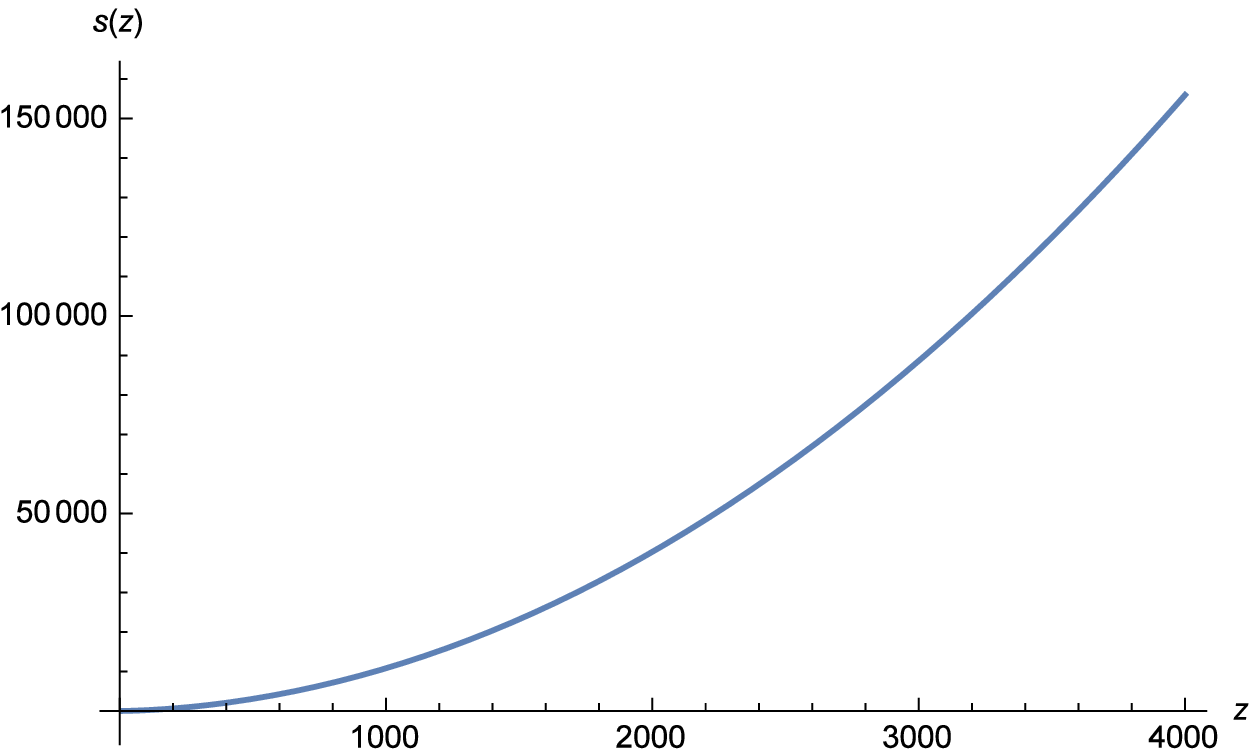}
\caption{Graphs of $s(z)$ for $z_{*} < z < z_{\rm tr}$ (left) 
and for $z_{\rm tr} < z < 4000$ (right).}
\label{sfull}
\end{figure}

We can entirely absorb the factors of $\alpha$ and $\Omega_{\rm c}$ in
equations (\ref{ztoZ}) and (\ref{exacteqn}) by changing the independent 
variable to $z$ and rescaling the dependent variables as,
\begin{equation}
f(z) \equiv -\frac{f_{y}(Z)}{\alpha^2 \Omega_{c}} \quad , \quad 
s(z) \equiv \frac{\sqrt{-Z}}{\alpha} \quad , \quad g(z) \equiv
\int_{z}^{\infty} \frac{f'(z') dz'}{(1 \!+\! z') s'(z') 
\widetilde{H}(z')} \; . \label{dimensionless}
\end{equation}
With these definitions equations (\ref{ztoZ}) and (\ref{exacteqn})
become,
\begin{eqnarray}
s(z) = (1 \!+\! z)^3 \!\! \int_{z}^{\infty} \!\! dz' \,
\frac{[\Omega_{\rm r} (1 \!+\! z')^4 + \frac12 \Omega_{\rm m} 
(1 \!+\! z')^3 - \Omega_{\Lambda}]}{(1 \!+\! z')^4 \, \widetilde{H}(z')} 
\; , \label{seqn} \\ 
\frac12 f(z) + \frac{\widetilde{H}(z) f'(z)}{2 s'(z)} + 
\widetilde{H}^2(z) g(z) = \frac1{12} (1 \!+\! z)^3 \; . \label{feqn}
\end{eqnarray}
Equation (\ref{feqn}) requires an initial condition which we take 
from the large $z$ limiting form of $f(z) \rightarrow \frac1{33}
(1 + z)^3$.

\begin{figure}[ht]
\includegraphics[width=6cm,height=4.8cm]{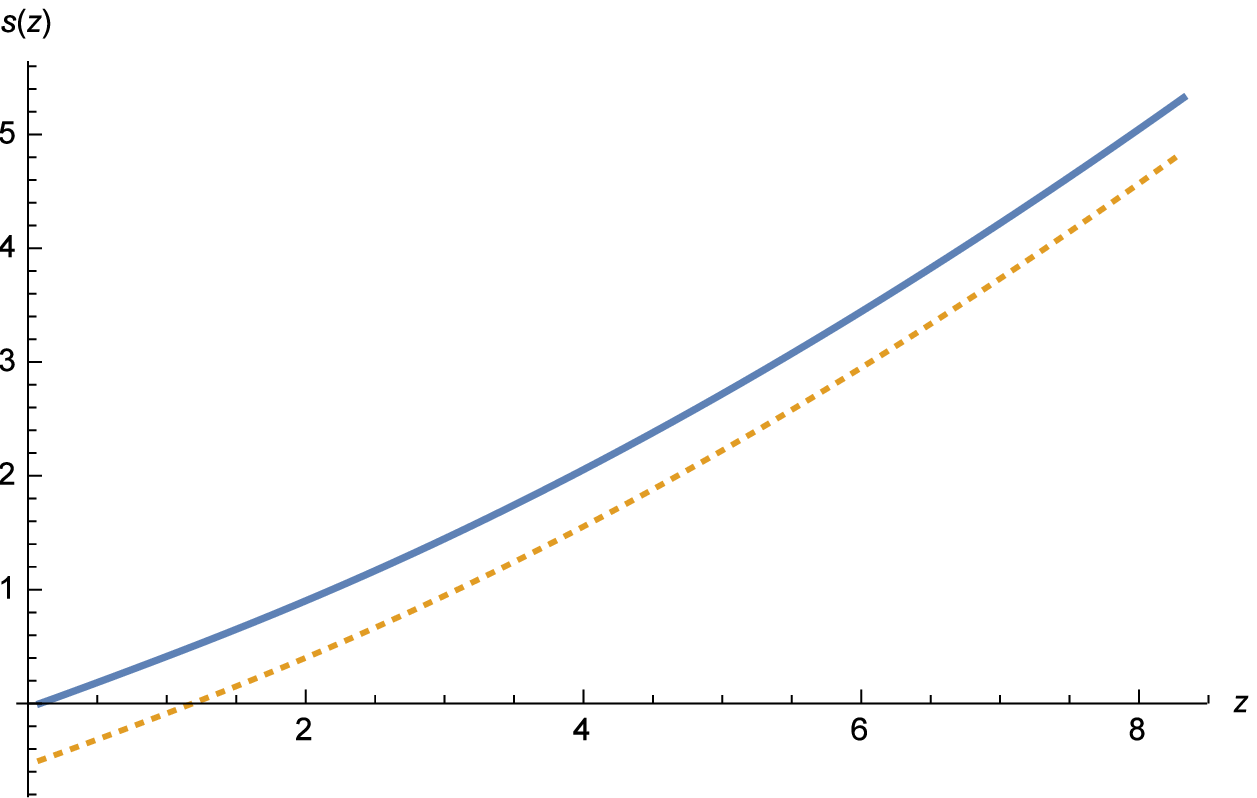}
\hspace{1cm}
\includegraphics[width=6cm,height=4.8cm]{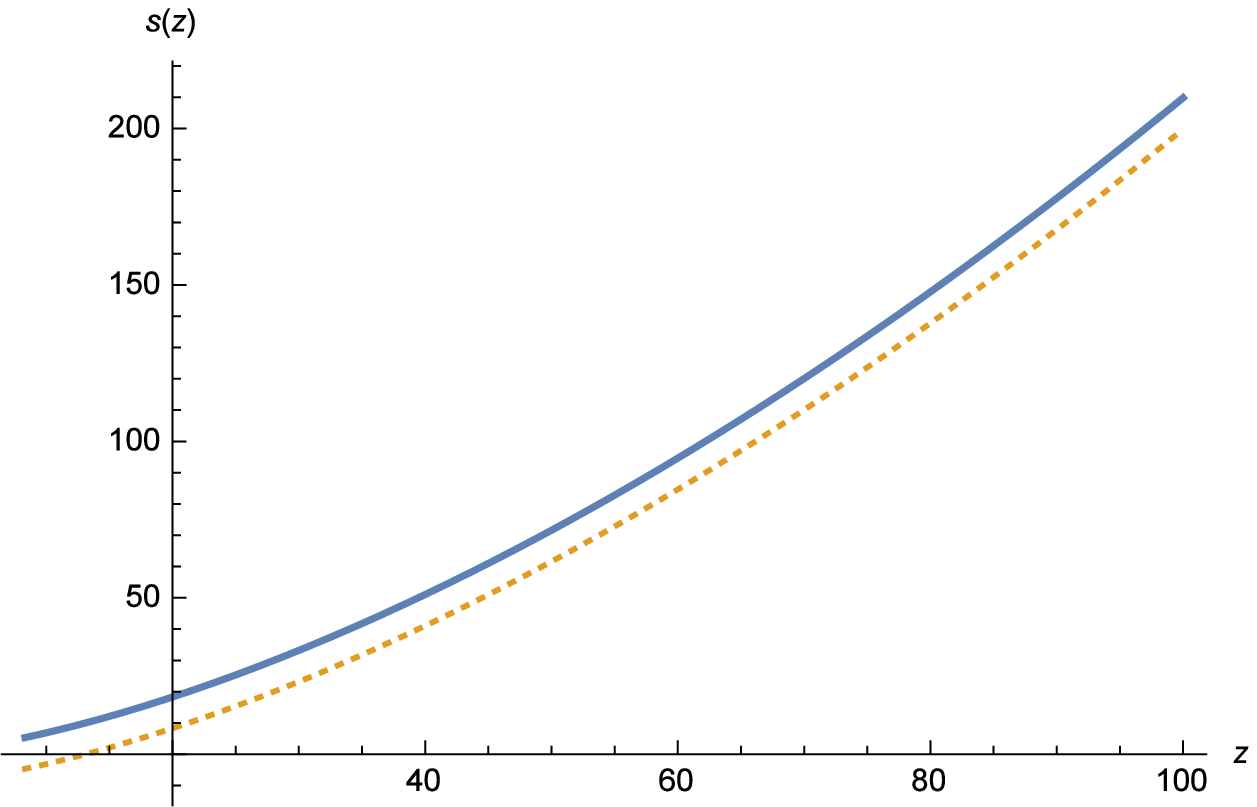}
\caption{Graphs of $s(z)$ for $z_{*} < z < z_{\rm tr}$ (left) 
and for $z_{\rm tr} < z < 100$ (right). The dotted line on
the left shows the analytic formula (\ref{ssmall}), minus 
an offset of $\Delta s = \frac12$ introduced to make the two 
curves distinguishable. The dotted line on the right displays 
the analytic formula (\ref{slarge}), minus an offset of $\Delta s 
= 10$ introduced to make the two curves distinguishable.}
\label{sanalytic}
\end{figure}

Figure~\ref{sfull} shows $s(z)$ over a large range of redshifts. 
Our previous work provided good analytic approximations for 
$s(z)$ --- expressions (\ref{Zall}) and (\ref{smallform}) --- 
depending upon whether $z$ is larger or smaller than the redshift
of radiation-vacuum equality $z_{\rm tr} \equiv 
(\frac{\Omega_{\Lambda}}{\Omega{\rm m}})^{\frac14} - 1 \simeq 8.32$,
\begin{eqnarray}
z > z_{\rm tr} & \!\!\!\! \Longrightarrow \!\!\!\! & 
s(z) = \frac{\sqrt{\Omega_{\rm r}}
(1 \!+\! z)^3}{1 \!+\! z_{\rm eq}} \Biggl\{ \frac13 \Bigl[1 +
\frac{1 \!+\! z_{\rm eq}}{1 \!+\! z}\Bigr]^{\frac32} + \Bigl[1 +
\frac{1 \!+\! z_{\rm eq}}{1 \!+\! z} \Bigr]^{\frac12} - \frac43
\Biggr\} , \qquad \label{slarge} \\
z < z_{\rm tr} & \!\!\!\! \Longrightarrow \!\!\!\! & s(z) = 
\frac{\sqrt{\Omega_{\Lambda}}}{3 y^2} \Biggl\{ (2 Y \!+\! 5 Y^3) 
\sqrt{1 \!+\! Y^2} - 4 Y^4 - 2 \ln\Bigl[Y \!+\! \sqrt{1 \!+\! Y^2} 
\, \Bigr] \nonumber \\
& & \hspace{5cm} - y \sqrt{1 \!+\! y^2} + 2 \ln\Bigl[y \!+\! 
\sqrt{1 \!+\! y^2} \, \Bigr] \Biggr\} . \qquad \label{ssmall}
\end{eqnarray}
Here the redshift of matter-vacuum equality is $z_{\Lambda} \equiv
(\frac{\Omega_{\Lambda}}{\Omega_{\rm m}})^{\frac13}$ and we define
$Y \equiv \sqrt{\frac{1 + z_{\rm tr}}{1 + z_{\rm eq}}} \simeq 0.0526$
and $y(z) \equiv (\frac{1 + z_{\Lambda}}{1 + z})^{\frac32}$.  
Figure~\ref{sanalytic} shows that expressions 
(\ref{slarge}-\ref{ssmall}) are in excellent agreement with the
exact result (\ref{seqn}). In fact, we had to offset the analytic
formulae in order to distinguish them from the exact result! The
same is not at all true for the asymptotic series expansions
(\ref{largeZ}) and (\ref{Ztoy}),
\begin{eqnarray}
{\rm Large}\, z & \!\!\!\! \Longrightarrow \!\!\!\! & s(z) =
\sqrt{\Omega_{\rm r}} \, (1 \!+\! z)^2 \Biggl\{ 1 + \frac1{24} x^2 
- \frac{1}{32} x^3 + O\Bigl( x^4\Bigr) \Biggr\} , 
\label{slargeasympt} \\
{\rm Small}\, z & \!\!\!\! \Longrightarrow \!\!\!\! & s(z) = \frac{
\sqrt{\Omega_{\Lambda}}}{3 y_{*}^2} \Biggl\{ -1.496 \!\times\! 
\Delta y - 1.036 \!\times\! \Delta y^2 + O\Bigl( \Delta y^3\Bigr)
\Biggr\} , \label{ssmallasympt}
\end{eqnarray}
where $x(z) \equiv \frac{1 + z_{\rm eq}}{1 + z}$, $y_* \equiv y(z_*)
\simeq 1.318$ and $\Delta y \equiv y - y_{*} < 0$. Figure~\ref{sseries}
shows that the large $z$ series (\ref{slargeasympt}) is only valid for 
$z \gtwid 2000$, and the small $z$ series (\ref{ssmallasympt}) is only
valid for $z \ltwid \frac12$.

\begin{figure}[ht]
\includegraphics[width=6cm,height=4.8cm]{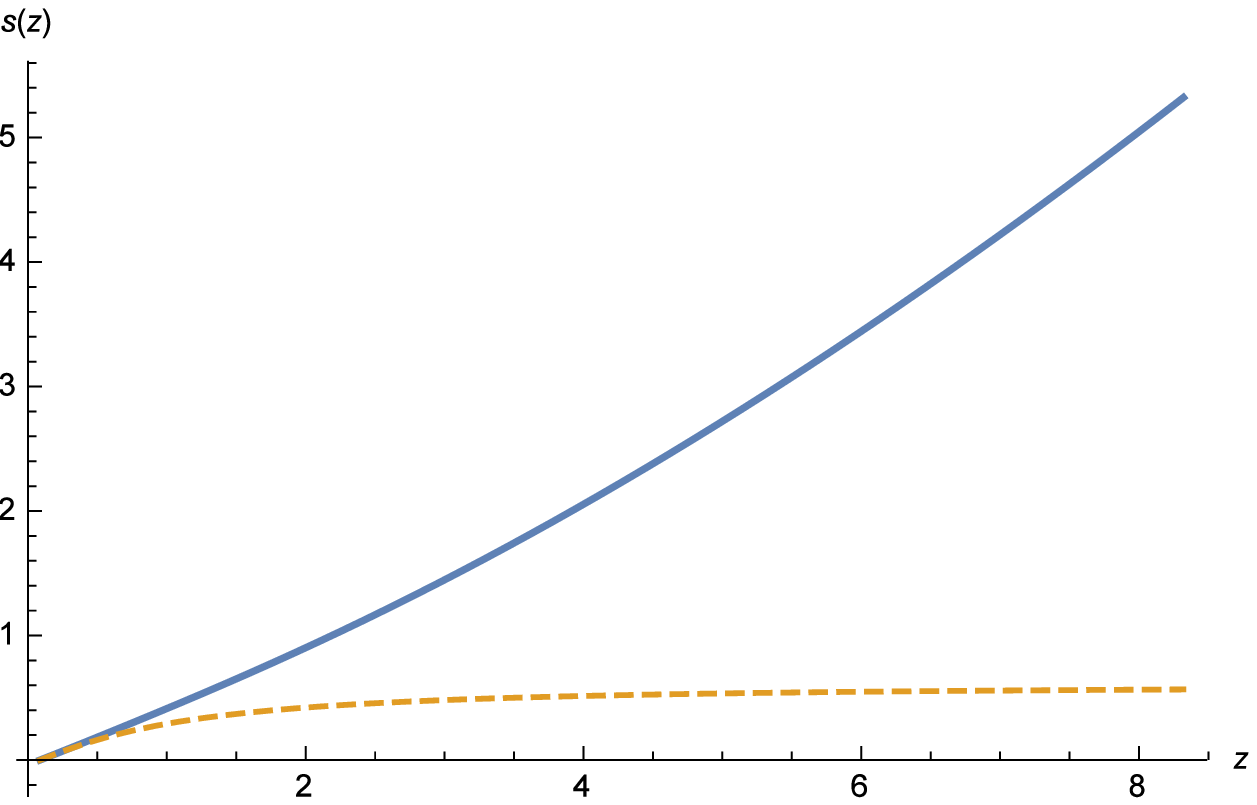}
\hspace{1cm}
\includegraphics[width=6cm,height=4.8cm]{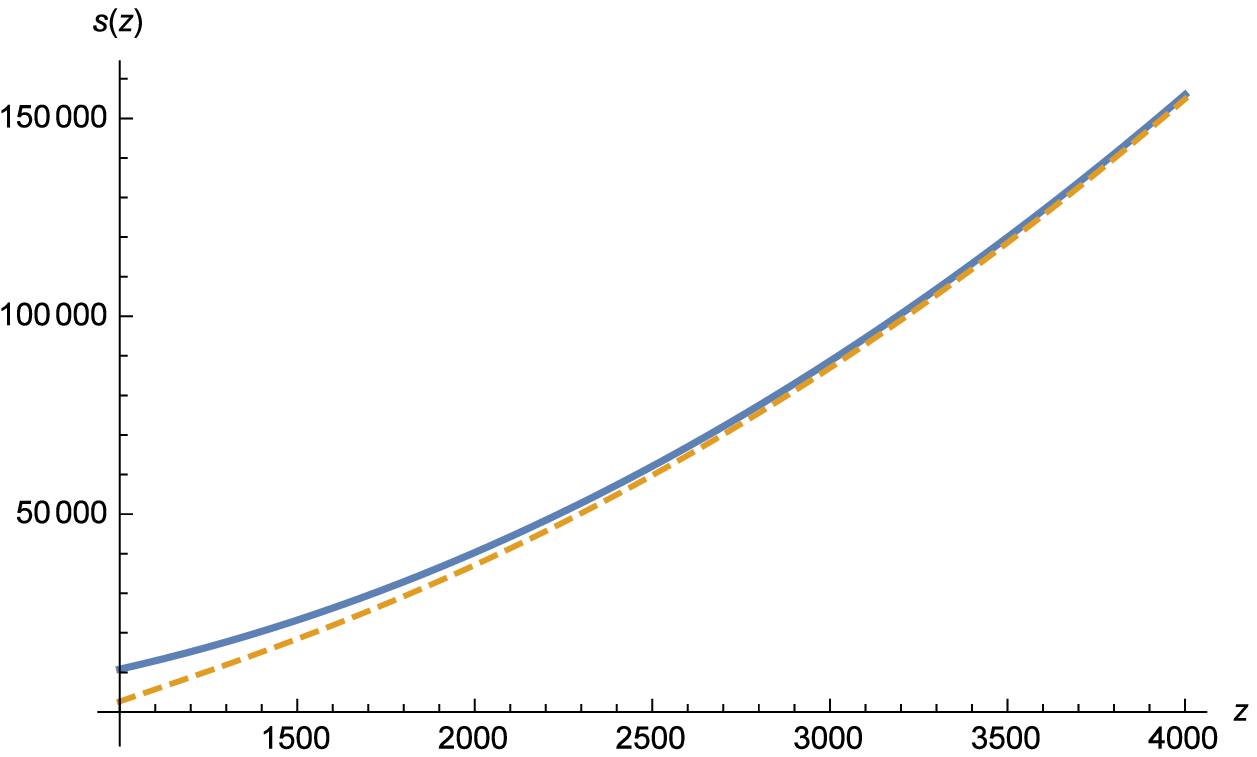}
\caption{Graphs of $s(z)$ for $z_{*} < z < z_{\rm tr}$ (left) 
and for $1000 < z < 4000$ (right). The dashed line on the left 
shows the small $z$ series expansion (\ref{ssmallasympt}) while 
the dashed line on the right shows the large $z$ series expansion 
(\ref{slargeasympt}).}
\label{sseries}
\end{figure}

\begin{figure}[ht]
\includegraphics[width=6cm,height=4.8cm]{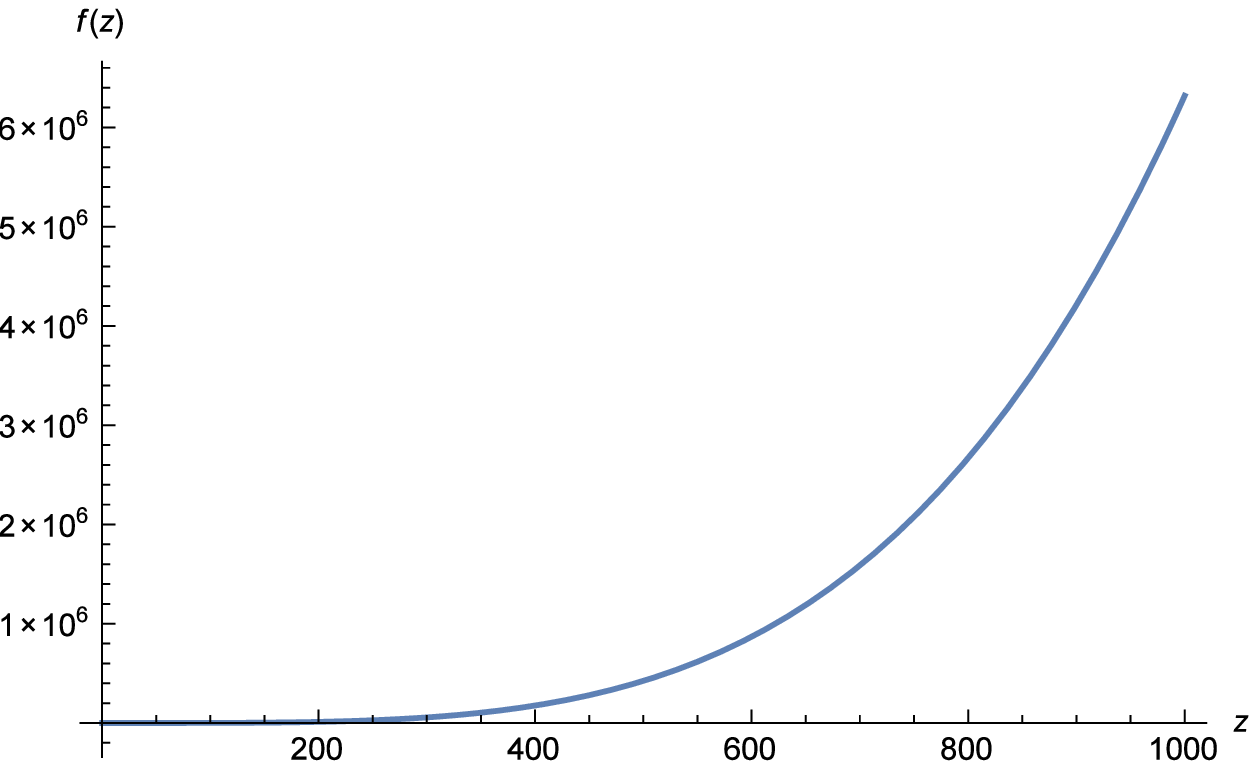}
\hspace{1cm}
\includegraphics[width=6cm,height=4.8cm]{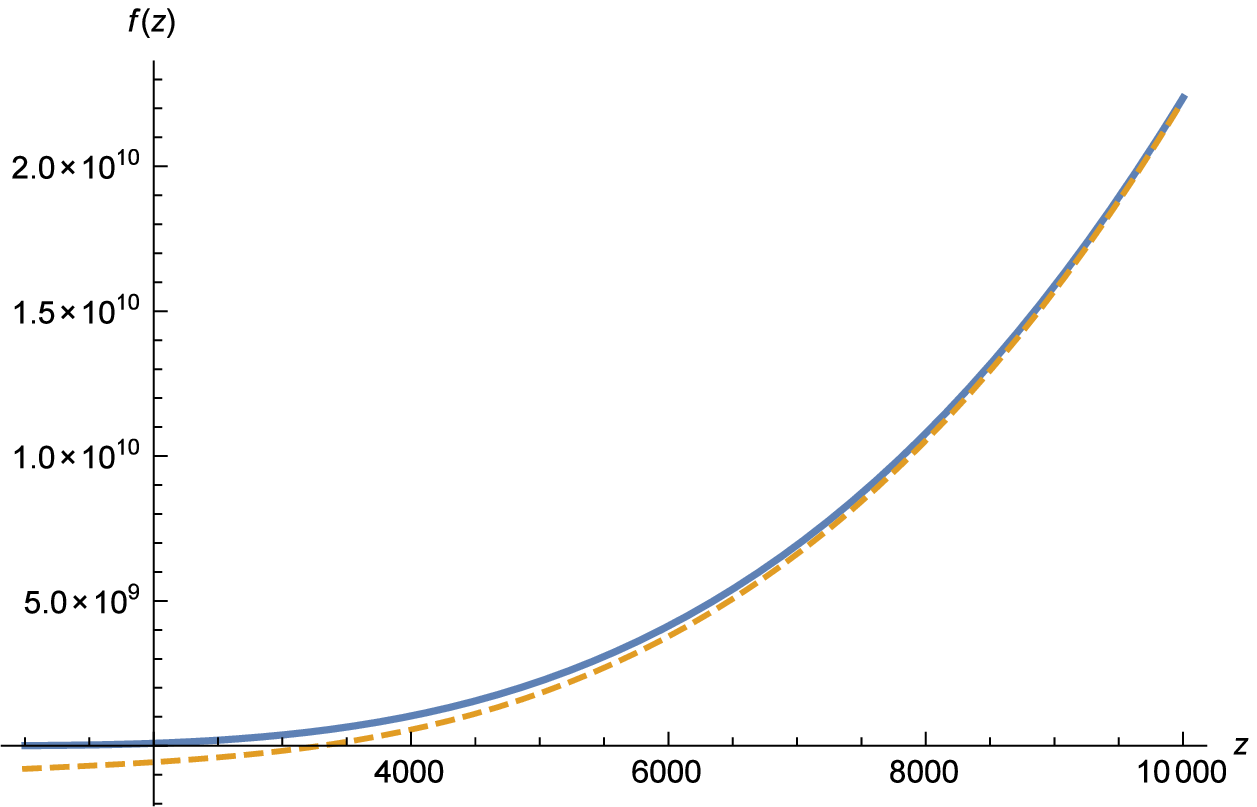}
\caption{Graphs of $f(z)$ for $z_{*} < z < 1000$ (left) 
and for $1000 < z < 10000$ (right). The dashed line on
the right-hand graph also shows the asymptotic series 
(\ref{largef}).}
\label{flarge}
\end{figure}

Evolving $f(z)$ with equation (\ref{feqn}) is more challenging
than numerically integrating $s(z)$ with (\ref{seqn}). First, there
is no analogue of the good analytic approximations 
(\ref{slarge}-\ref{ssmall}) we were able to get for $s(z)$. Our 
previous work --- expressions (\ref{largeZansatz}) and 
(\ref{smallZansatz}) --- does imply asymptotic series expansions for 
large and small $z$,
\begin{eqnarray}
{\rm Large}\, z & \!\!\!\! \Longrightarrow \!\!\!\! & f(z) = 
\frac{(1 \!+\! z)^3}{33} \Biggl\{ 1 - x + \frac{83}{88} x^2 - 
\frac{231}{256} x^3 + O\Bigl(x^4\Bigr) \Biggr\} , \label{largef} \\
{\rm Small}\, z & \!\!\!\! \Longrightarrow \!\!\!\! & f(z) = 
\frac{(1 \!+\! z_{\Lambda})^3}{36 y_{*}^2} \Biggl\{ 1.143 \!\times\!
\Delta y - 0.7373 \!\times\! \Delta y^2 + O\Bigl( \Delta y^3\Bigr)
\Biggr\} . \qquad \label{smallf}
\end{eqnarray}
However, one can see from Figures~\ref{flarge} and \ref{fsmallmid}
that these series approximations are only accurate for $z \gtwid 
5000$ and for $z \ltwid 1$, respectively. 

\begin{figure}[ht]
\includegraphics[width=6cm,height=4.8cm]{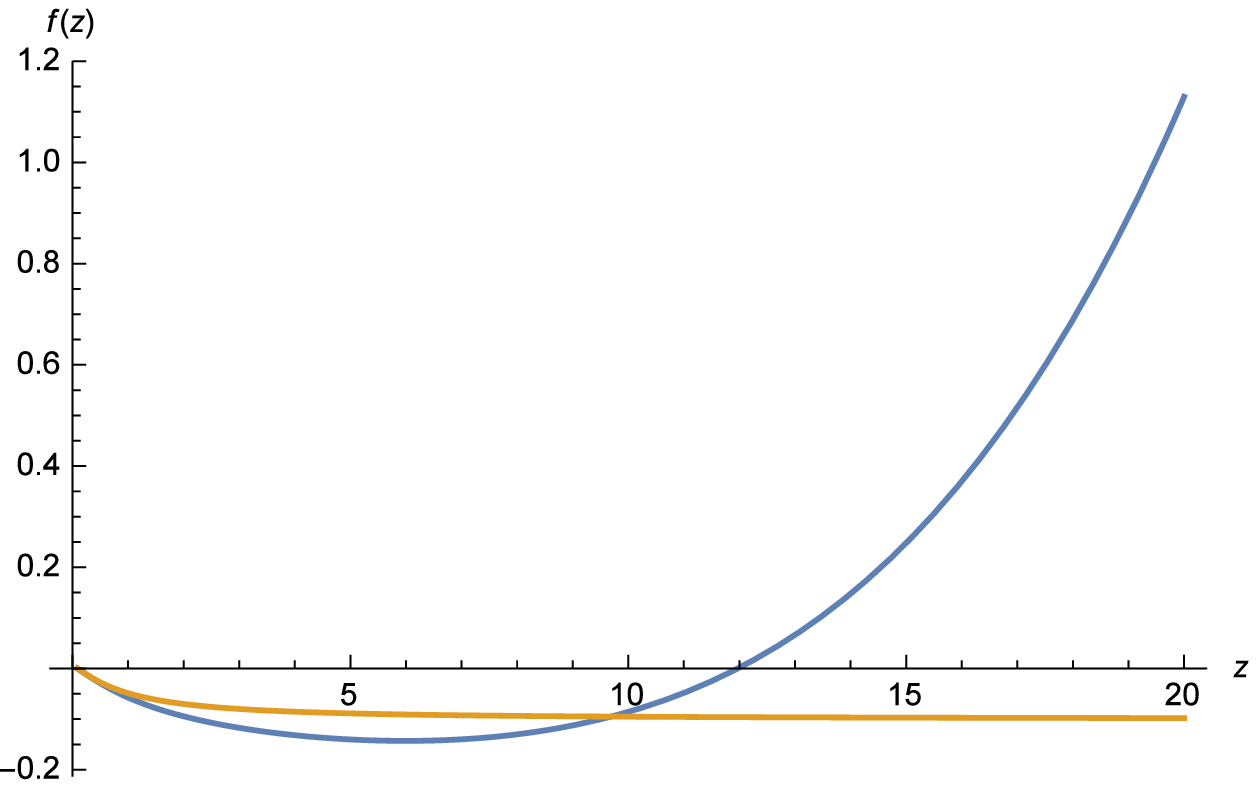}
\hspace{1cm}
\includegraphics[width=6cm,height=4.8cm]{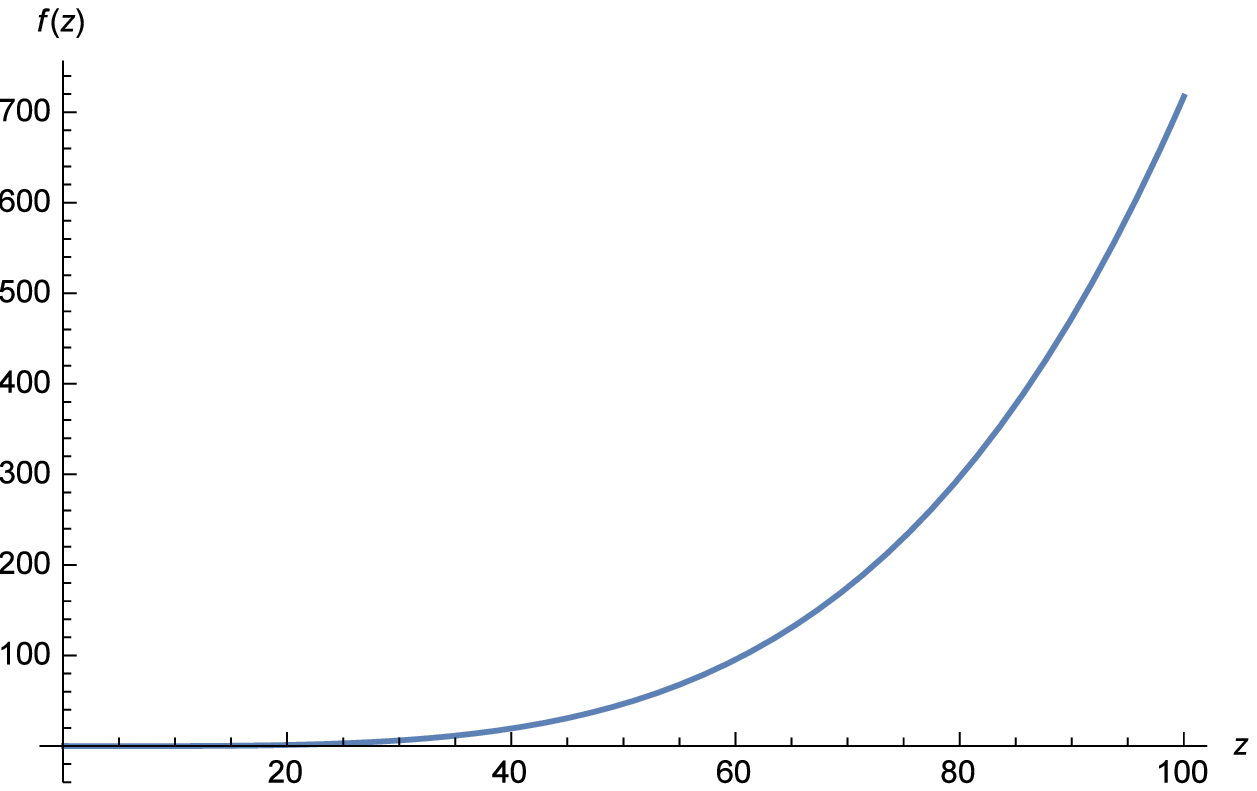}
\caption{Graphs of $f(z)$ for $z_{*} < z < 20$ (left) 
and for $z_{*} < z < 100$ (right). The dashed line on the left
also shows the asymptotic series (\ref{smallf}).}
\label{fsmallmid}
\end{figure}

A worse problem is that numerical solution of (\ref{feqn}) is unstable. 
We can evolve from large $z$ to small, starting from the excellent 
series approximation (\ref{largef}), but the result tends to diverge 
for small $z$. The better strategy turns out to be evolving from small 
$z$ to large, starting from $f(z_{*}) = 0$. To facilitate this procedure 
one must extract $g_{*} \equiv g(z_{*})$ from $g(z)$,
\begin{equation}
g(z) = g_{*} - \int_{z_{*}}^{z} \!\! \frac{f'(z') dz'}{(1 \!+\! z') 
s'(z') \widetilde{H}(z')} \; . \label{newg}
\end{equation}
When this is done, evolving to arbitrarily large $z$ produces a 
solution which seems to reach the form (\ref{largef}), but then grows
in magnitude like $z^{2 + 2 \sqrt{3}}$, either in the positive or 
negative direction. The correct value of $g_{*} \simeq 0.1807$ is found 
by seeking the point at which the asymptotic form changes sign.    

\subsection{Solving for $f_y(Z)$ Numerically}

\begin{figure}[ht]
\includegraphics[width=6cm,height=4.8cm]{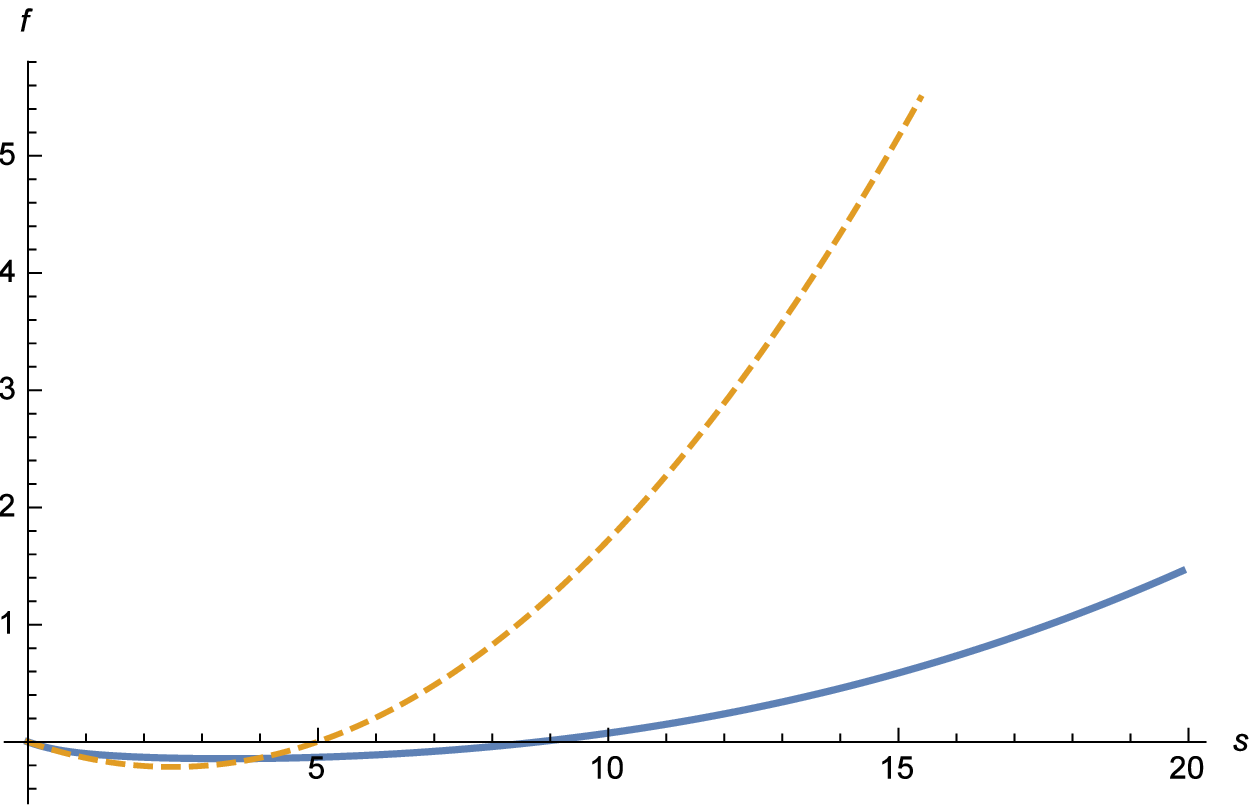}
\hspace{1cm}
\includegraphics[width=6cm,height=4.8cm]{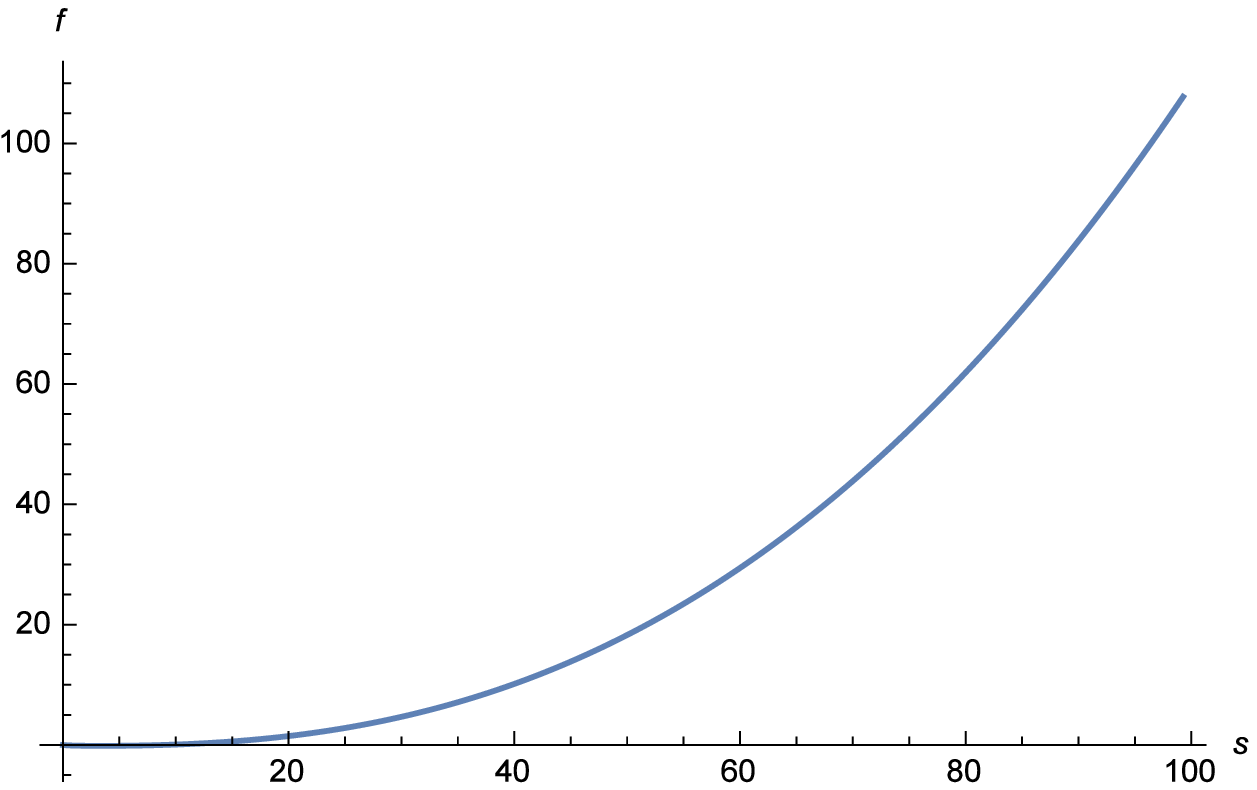}
\caption{Graphs of $f(z)$ versus $s$ for $0 < s < 20$ 
(left) and for $0 < s < 100$ (right). The dashed line on the left
gives the asymptotic series expansion (\ref{smalls}).}
\label{f(s)smallmid}
\end{figure}

\begin{figure}[ht]
\includegraphics[width=6cm,height=4.8cm]{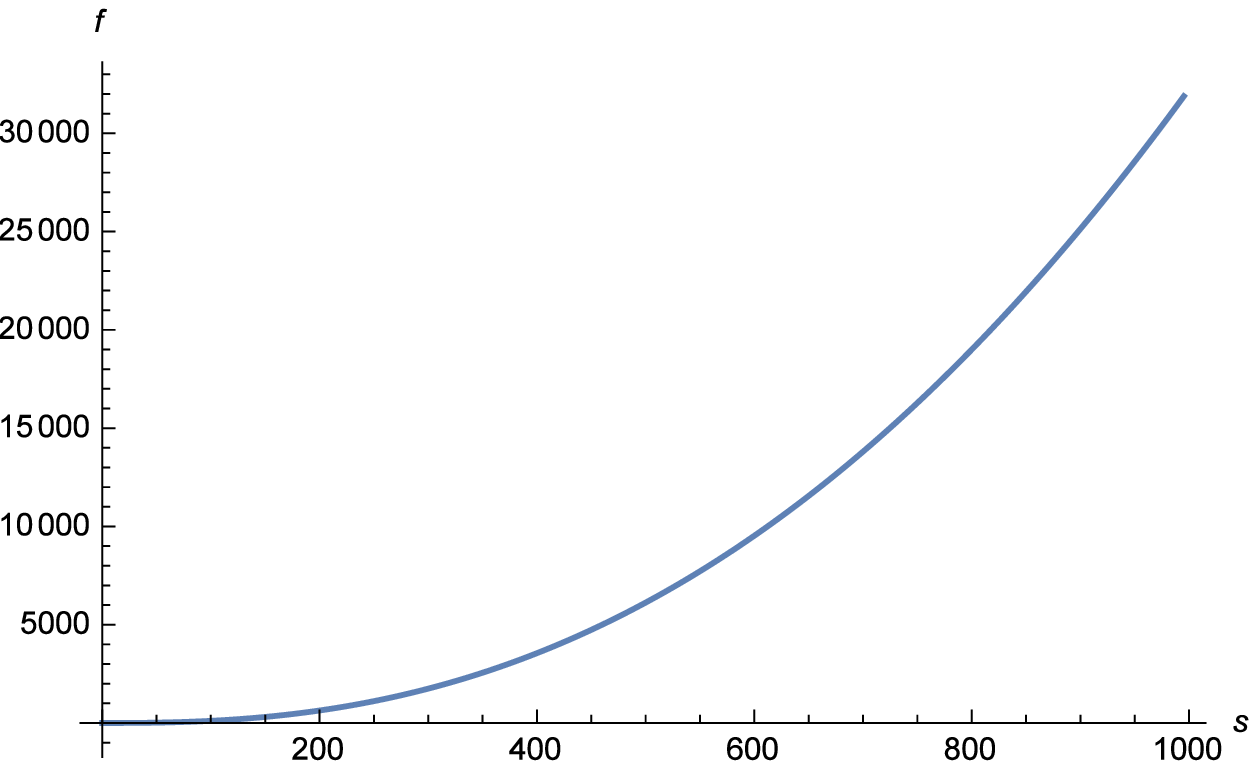}
\hspace{1cm}
\includegraphics[width=6cm,height=4.8cm]{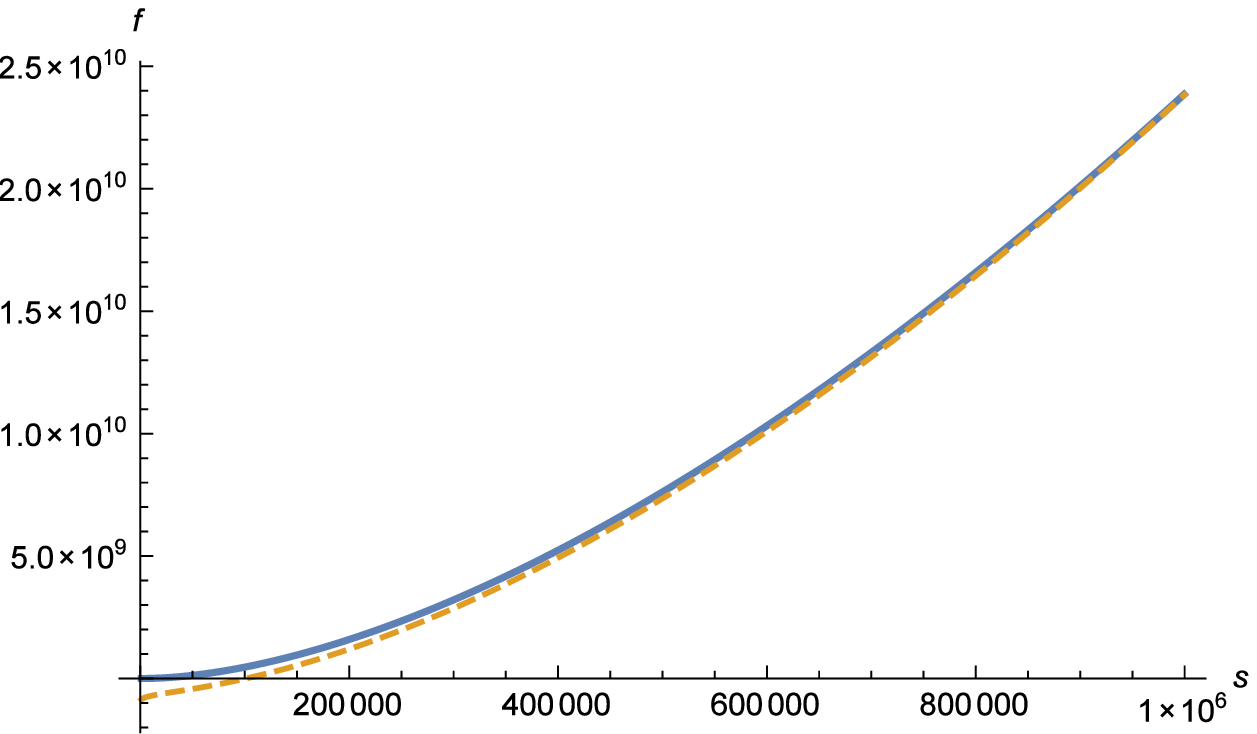}
\caption{Graphs of $f(z)$ versus $s$ for $0 < s < 10^3$ 
(left) and for $10^5 < s < 10^6$ (right). The dashed line 
on the right gives the asymptotic series expansion 
(\ref{larges}).}
\label{f(s)largeasympt}
\end{figure}

This is really just a matter of using our previous results for $s(z)$
and $f(z)$ to plot $f(z)$ as a function of $s$. Figures~\ref{f(s)smallmid}
and \ref{f(s)largeasympt} show the results for small and large values of
$s$, respectively. For the smallest ($0 < s < 20$) and largest 
($10^3 < s < 10^6$) ranges we also show the comparison with the
analytic asymptotic series expansion which follow from the work of
section 3,
\begin{eqnarray}
{\rm Small} \; s & \Longrightarrow & f(z) = \frac{(1 \!+\! 
z_{\Lambda})^3}{12} \Biggl\{ \frac{A \, s}{\sqrt{\Omega_{\Lambda}}} +
\frac{B \, s^2}{\Omega_{\Lambda}} + O\Bigl( \frac{s^3}{\Omega_{\Lambda}^{
\frac32}} \Bigr) \Biggr\} \label{f(smalls)} \qquad , \label{smalls} \\
{\rm Large} \; s & \!\!\! \Longrightarrow \!\!\! & f(z) = \frac1{33} 
\Bigl( \frac{s}{\sqrt{\Omega_{\rm r}}}\Bigr)^{\frac32} \! \Biggl\{\! 1 
\!-\! \frac1{\sqrt{\sigma}} \!+\! \frac{155}{176} \frac1{\sigma} \!-\! 
\frac{625}{768} \frac1{\sigma^{\frac32}} \!+\! O\Bigl( \frac1{\sigma^2}
\Bigr) \! \Biggr\} , \qquad \label{larges}
\end{eqnarray}
where $A \simeq -0.764$ and $B \simeq +0.127$ and we define $\sigma(s)$ as,
\begin{equation}
\sigma(s) \equiv \frac{s}{\sqrt{\Omega_{\rm r}} (1 \!+\! z_{\rm eq})^2} 
\; . \label{sigmadef}
\end{equation}
Points to note are that the small $s$ series (\ref{smalls}) breaks down
for $s \gtwid \frac14$, and the large $s$ series (\ref{larges}) breaks
down for $s \ltwid 300,000$.

\subsection{Fitting $f_{y}(Z)$ to an Analytic Form}

\begin{figure}[ht]
\includegraphics[width=6cm,height=4.8cm]{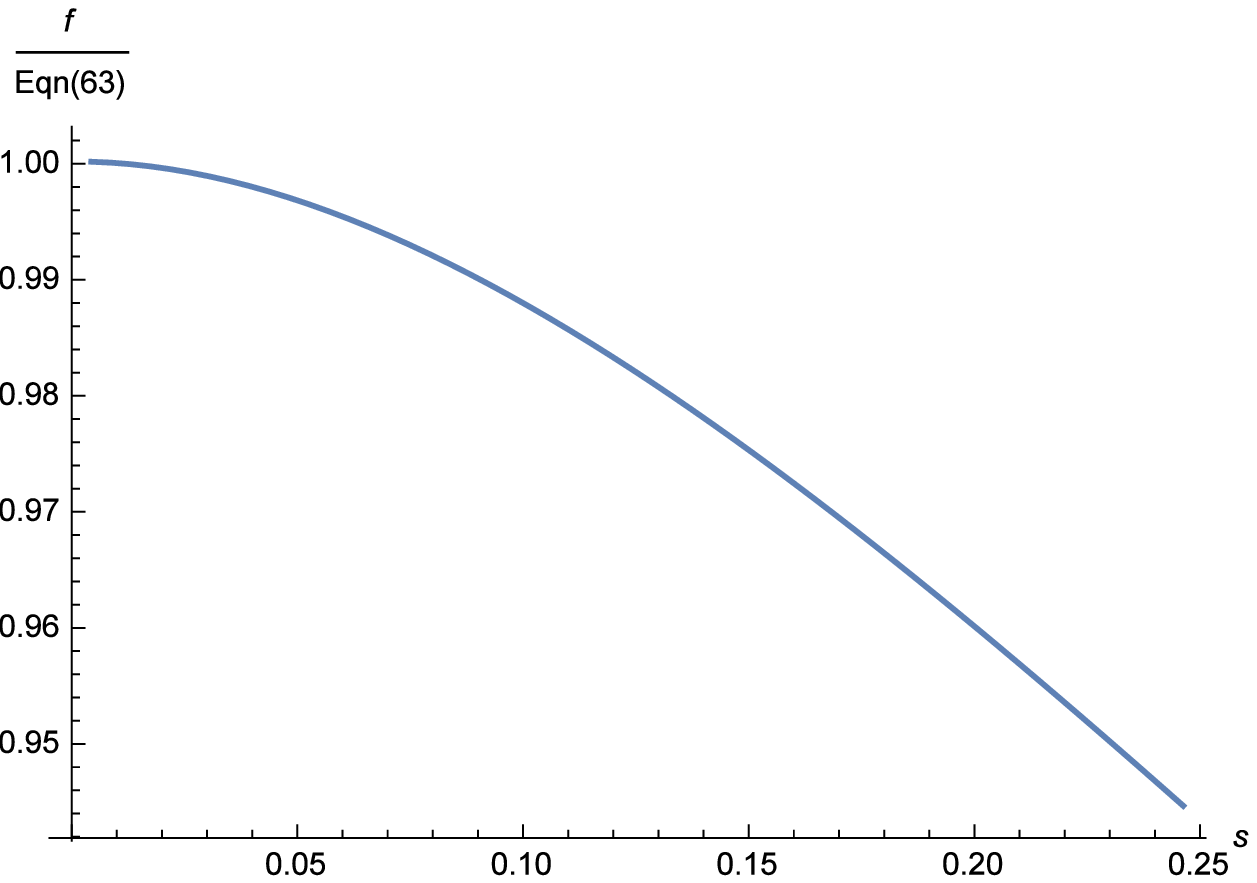}
\hspace{1cm}
\includegraphics[width=6cm,height=4.8cm]{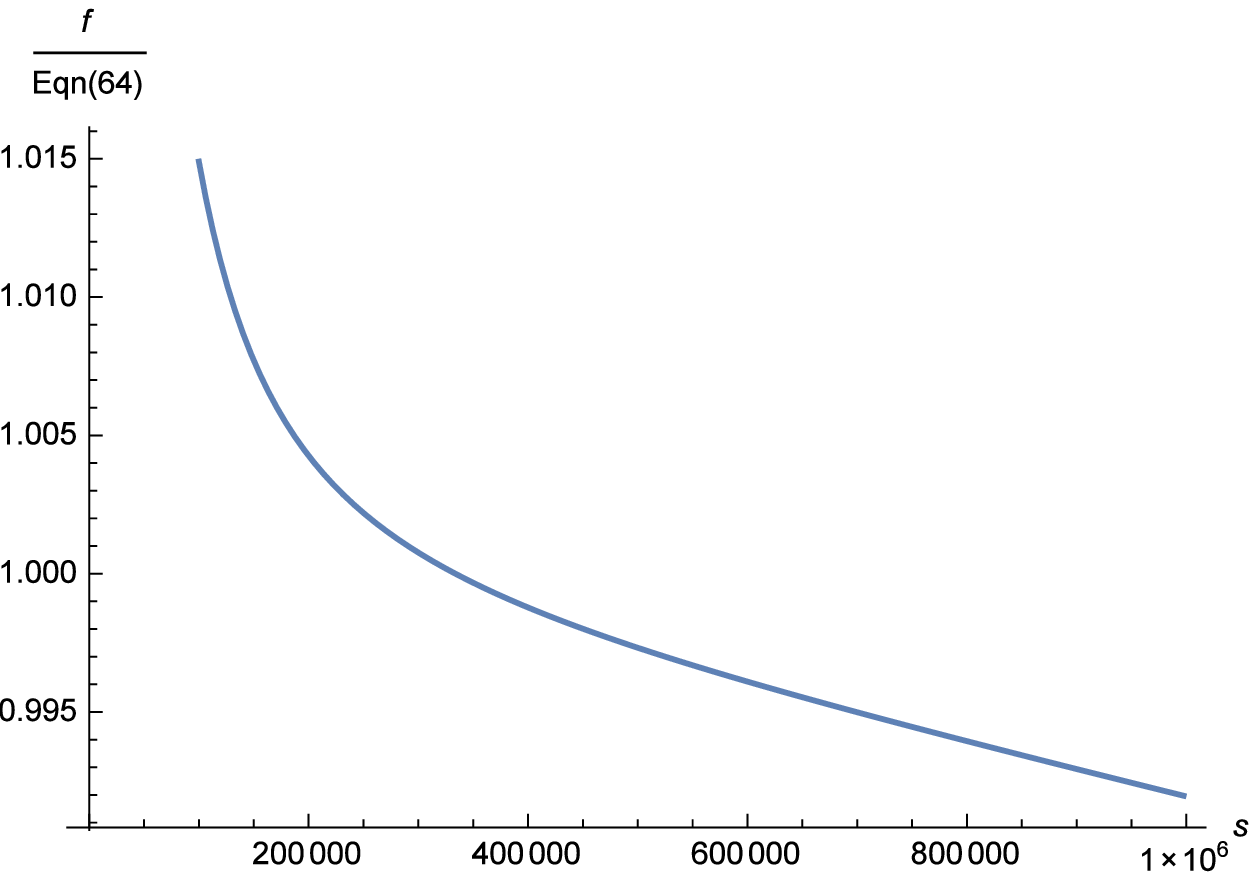}
\caption{Graphs of the ratios of $f(z)$ to the asymptotic series
expansions (\ref{smalls}-\ref{larges}) versus $s$ for the ranges 
$0 < s < \frac14$ (on the left) and  for $10^5 < s < 10^6$ (on the
right).}
\label{ratios}
\end{figure}

A better measure of the accuracy of the asymptotic series expansions 
(\ref{smalls}-\ref{larges}) is gained by plotting the ratio of the 
numerical result for $f(z)$ divided by the expansions. Figure~\ref{ratios}
shows this. The failure of the very large $s$ ratio to exactly approach 
unity is due to instability of our numerical determination of $f(z)$, as
explained before. However, the deviations for $\frac14 < s < 300,000$ 
represent the transition between the two asymptotic forms which we would
like to fit to an analytic function.

\begin{figure}[ht]
\includegraphics[width=6cm,height=4.8cm]{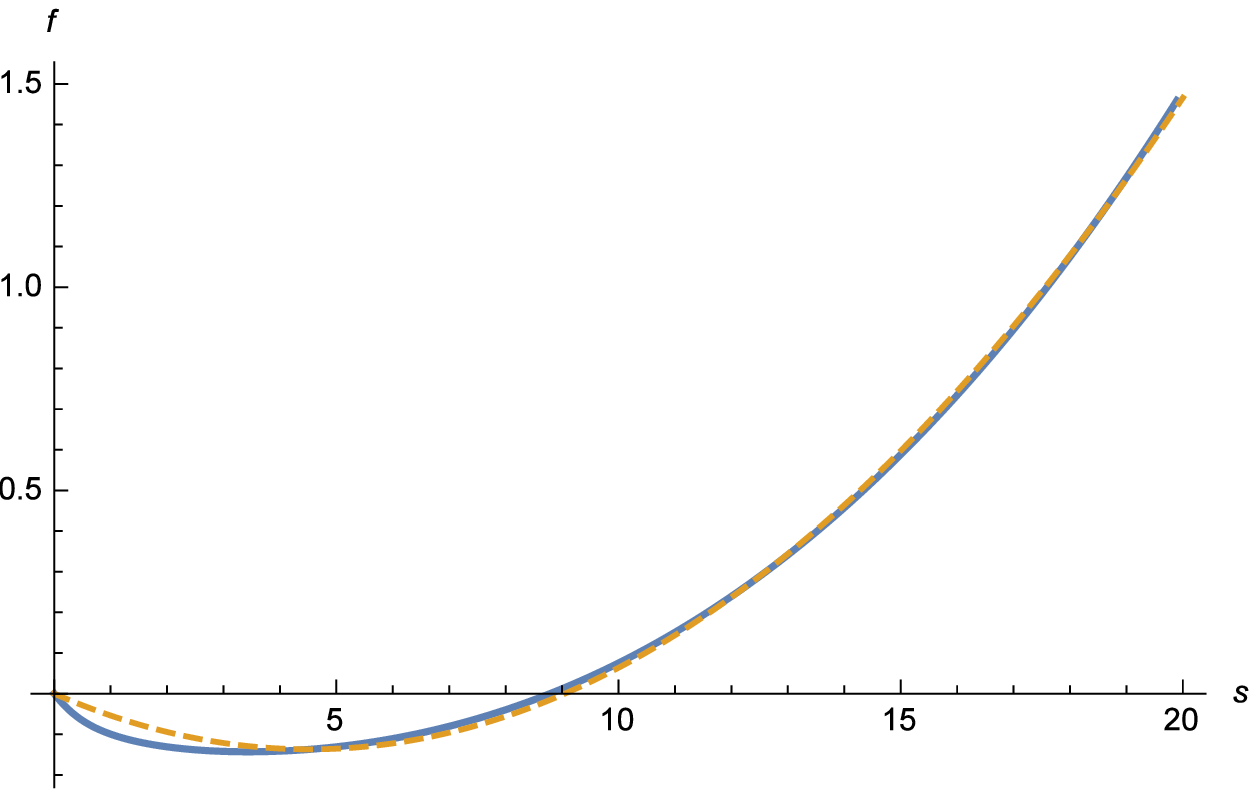}
\hspace{1cm}
\includegraphics[width=6cm,height=4.8cm]{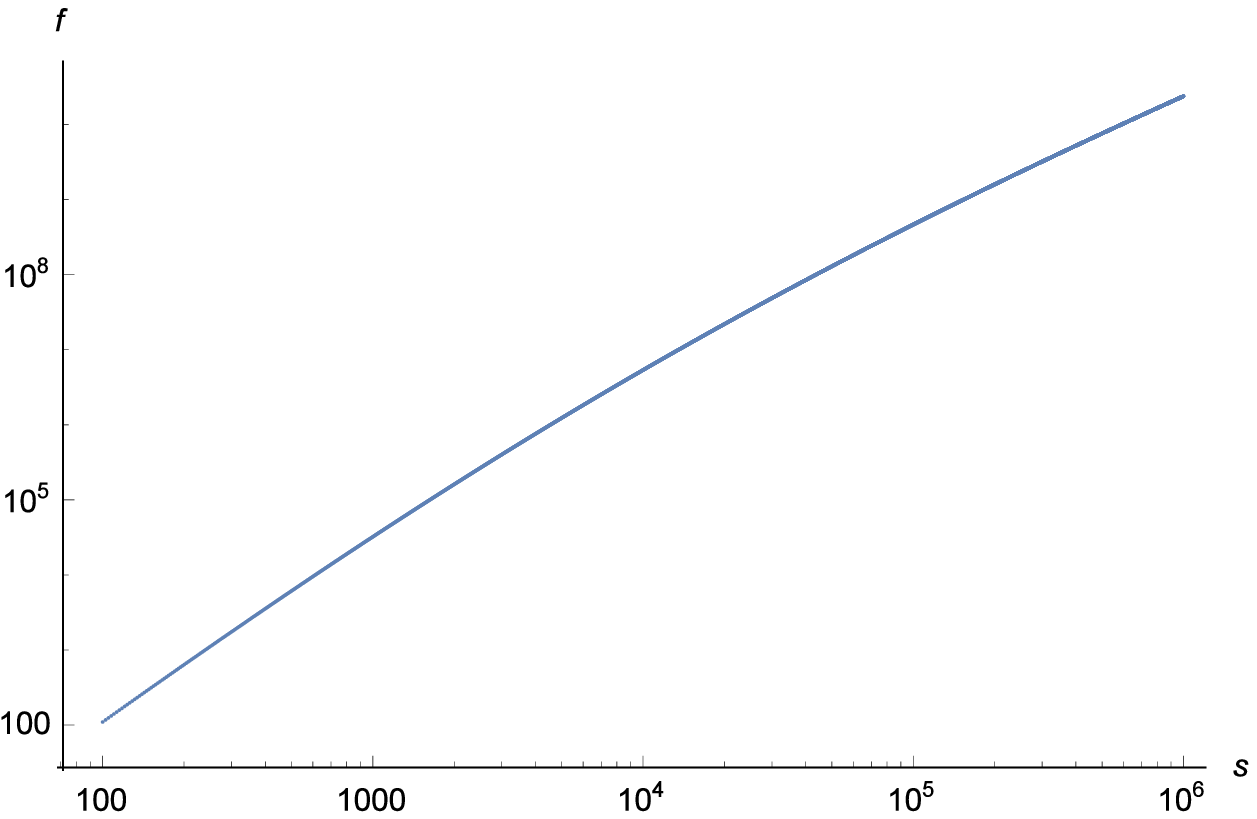}
\caption{The left hand graph shows $f(z)$ versus $s(z)$ in the region
$0 < s < 20$. The dashed line shows the best fit quadratic in this
same region. The right hand graph is a log-log plot of $f(z)$ versus
$s$ in the range $10^2 < s < 10^6$.}
\label{qualitative}
\end{figure}

\begin{figure}[ht]
\includegraphics[width=6cm,height=4.8cm]{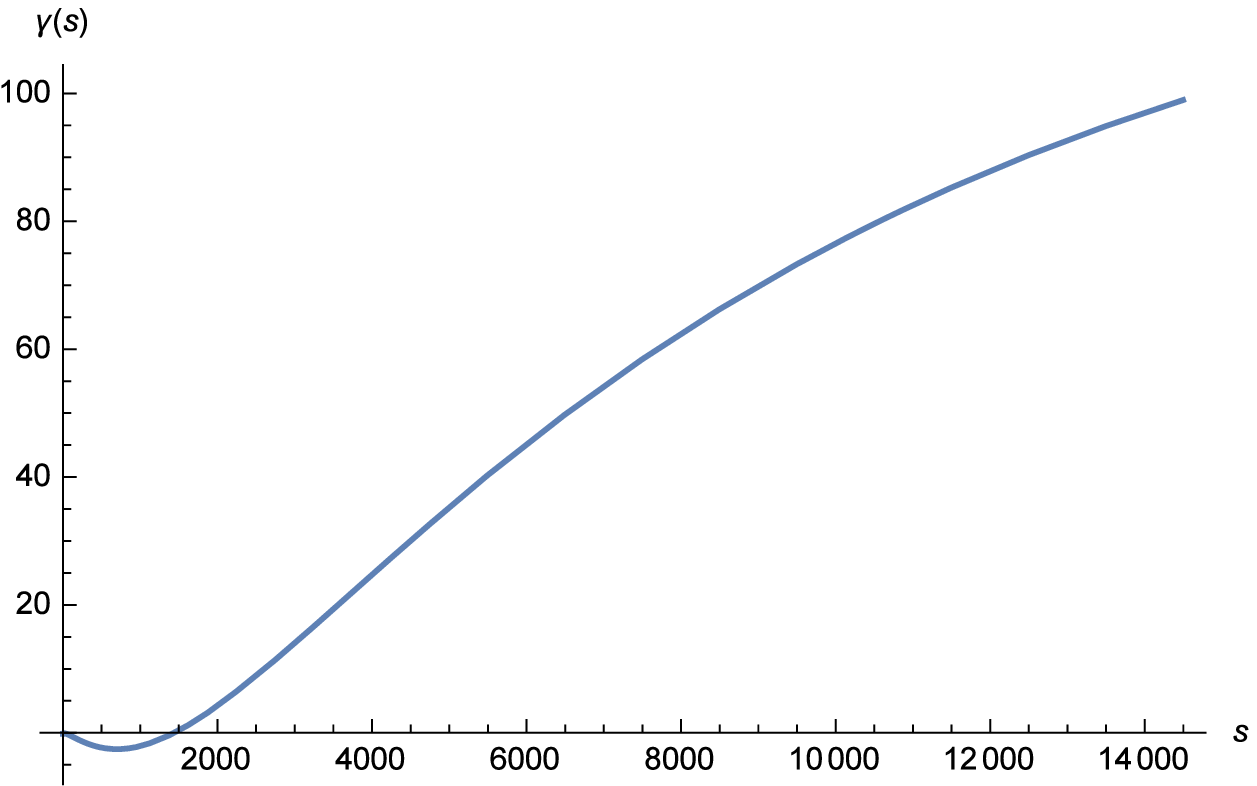}
\hspace{1cm}
\includegraphics[width=6cm,height=4.8cm]{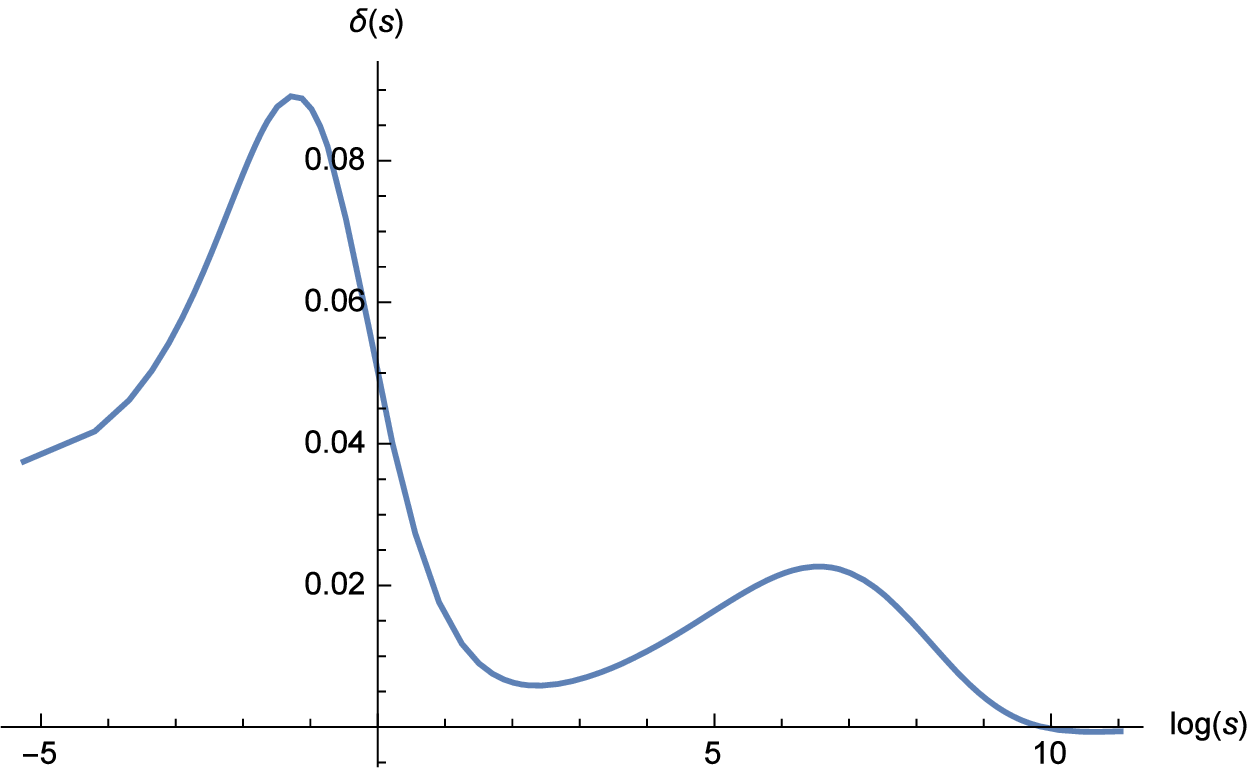}
\caption{Variation of the coefficients $\gamma(s)$ and $\delta(s)$ in 
equation (\ref{ansatz}). The left hand graph shows how the fitted 
coefficient $\gamma$ varies slowly with the midpoint of a small region 
used to make the fit. The right hand graph shows the same thing for the 
coefficient $\delta$.}
\label{variation}
\end{figure}

We are dealing with a smooth function whose qualitative features are
shown in Figure~\ref{qualitative}:
\begin{itemize}
\item{The small $s$ regime is well fit by $\gamma s + \delta s^2$, but
the coefficients $\gamma$ and $\delta$ evolve slowly as shown in 
Figure~\ref{variation}; and}
\item{The large $s$ regime is well fit by a power law which changes 
slowly from $s^{2}$ for moderate values of $s$ to $s^{\frac32}$ for 
large values of $s$.}
\end{itemize}
We can find a reasonable ansatz by making a plausible interpolation 
of the asymptotic series in curly brackets of expression (\ref{larges}),
with $\sigma(s)$ from (\ref{sigmadef}),
\begin{equation}
1 - \frac1{\sqrt{\sigma}} + \frac{155}{176} \frac1{\sigma} - \frac{625}{768} \frac1{\sigma^{\frac32}} + O\Bigl( \frac1{\sigma^2} \Bigr) \longrightarrow 
\frac1{1 + \frac1{\sqrt{\sigma}} + \frac{21}{176 \, \sigma} + 
\frac{443}{8448 \, \sigma^{\frac32}} } \; . \qquad \label{interpolation}
\end{equation}
The right hand side of (\ref{interpolation}) has the same large $\sigma$ 
series expansion as the left hand side, but it behaves like $\frac{8448}{443}
\sigma^{\frac32}$ for small $\sigma$. Hence we might define,
\begin{equation}
f_{\rm lg}(s) \equiv \frac1{33} \Bigl( \frac{s}{\Omega_{\rm r}}\Bigr)^{\frac32}
\!\times\! \frac1{1 + \frac1{\sqrt{\sigma(s)}} +\frac{21}{176 \, \sigma(s)} + 
\frac{443}{8448 \, \sigma^{\frac32}(s)} } \; . \label{goodlarge}
\end{equation}
This function will recover the asymptotic large $s$ behavior, and go to zero
rapidly for small $s$. Then a reasonable ansatz for the full function is,
\begin{equation}
f(z) = \gamma(s) \!\times\! s + \delta(s) \!\times\! s^2 + f_{\rm lg}(s) \; ,
\label{ansatz}
\end{equation}

The general shapes of $\gamma(s)$ and $\delta(s)$ in Figure~\ref{variation}
motivate our fits. For $\gamma(s)$ a single rational function suffices; 
$\delta(s)$ is better described by the sum of two terms, each of which falls 
off exponentially. Our results are,
\begin{eqnarray}
\gamma(s) & = & \frac{(\frac5{29} \!+\! \frac{s}{144}) (\frac{s}{1440} \!-\! 
1)}{1 + \frac{s}{10000} + (\frac{s}{5700})^2 + (\frac{s}{25000})^3} \; , 
\label{gamma} \\
\delta(s) & = & \frac{\frac1{29} \!+\! \frac19 \sqrt{s} \!-\! \frac16 s \!+\! 
\frac29 s^{\frac32} + s^2}{1 + 25 s^{\frac{11}{4}} \exp[\frac1{12} s]} +
\frac{ \frac{s}{4000} \!-\! 100 (\frac{s}{4000})^2 \!+\! 10^4 
(\frac{s}{4000})^3}{1 + 33 \!\times\! 10^4 (\frac{s}{4000})^{\frac{32}{11}} 
\exp[\frac{s}{4000}]} \; . \label{delta}
\end{eqnarray}
More complicated polynomials would of course be more accurate.

\subsection{Quantifying the Disagreement for $0 < z < z_{*}$}

\begin{figure}[ht]
\includegraphics[width=6cm,height=4.8cm]{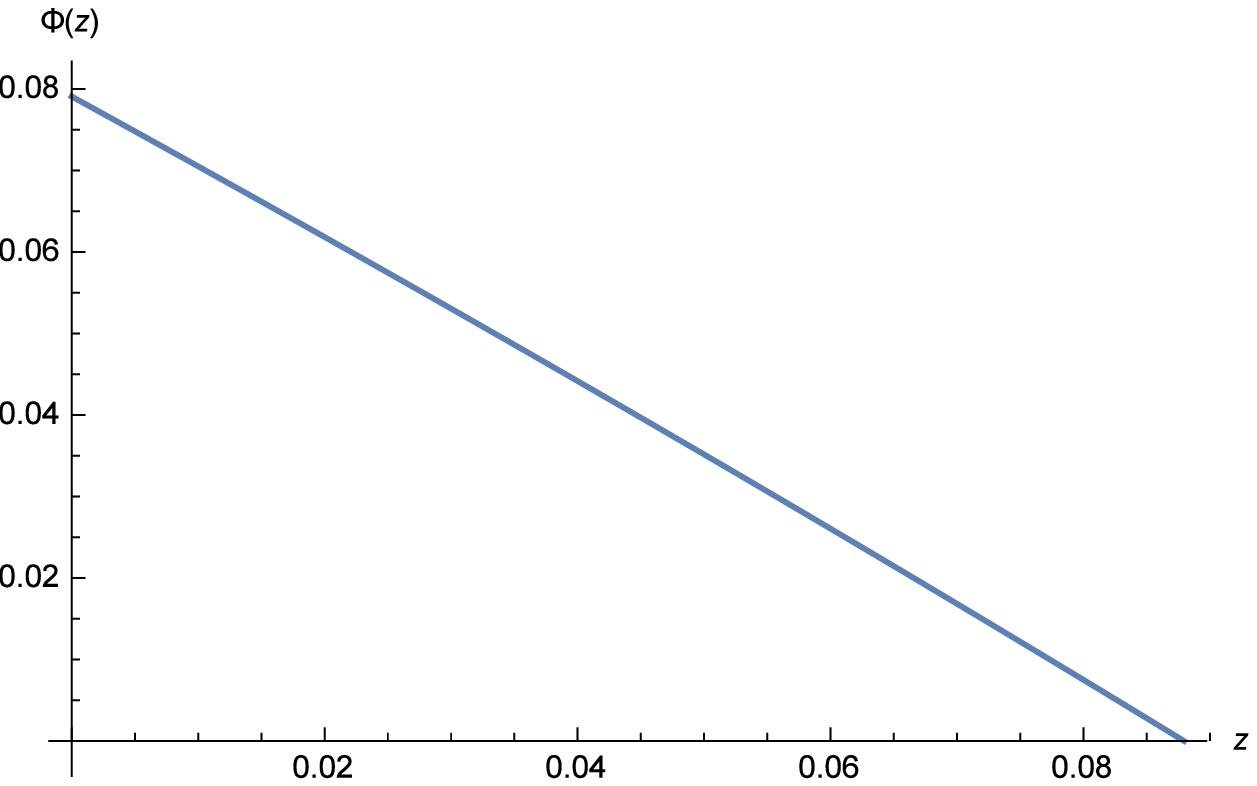}
\hspace{1cm}
\includegraphics[width=6cm,height=4.8cm]{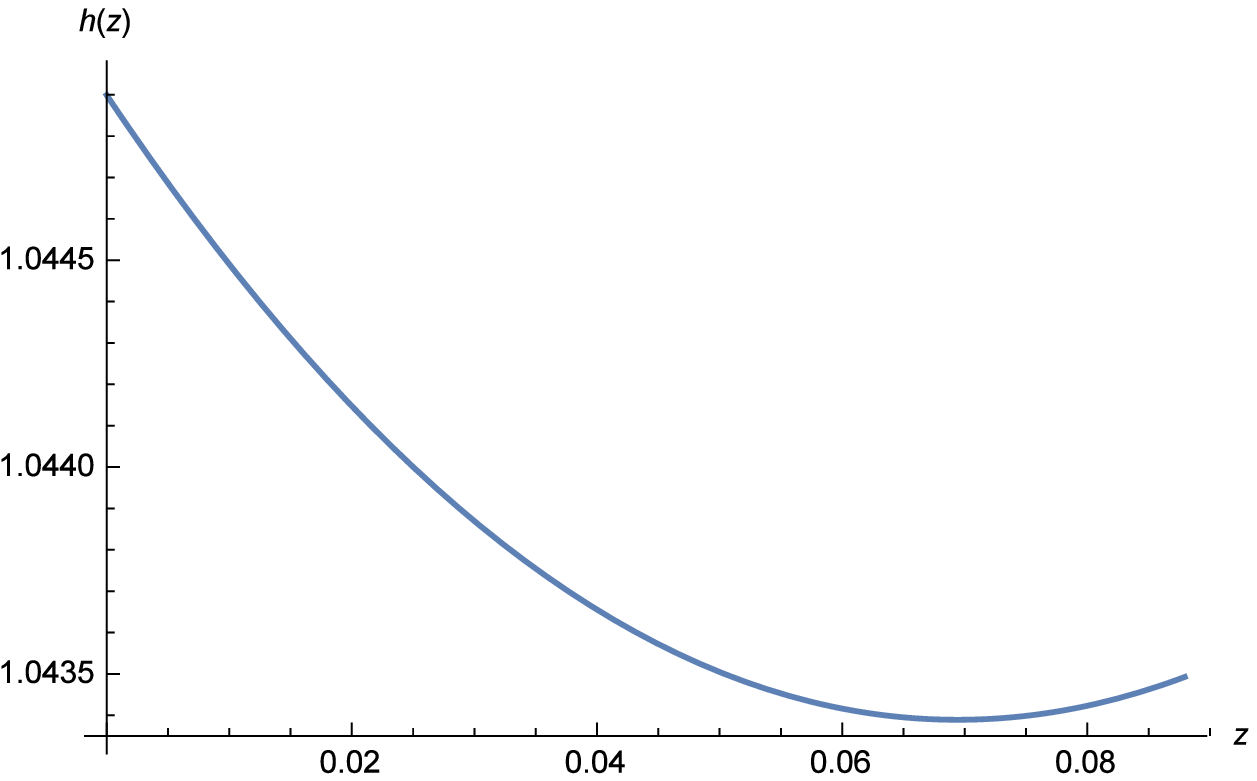}
\caption{The left hand graph shows the evolution of $\Phi(z) \equiv 
\frac{\dot{\phi}}{6 H_0}$ in the actual model. The right hand graph
shows $h(z) \equiv \frac{H}{H_0}$ in the actual model.}
\label{modelresults}
\end{figure}

We have shown how to choose the function $f_{y}(Z)$ so as to exactly reproduce 
the $\Lambda$CDM expansion history for all redshifts greater than $z_{*} \simeq 
0.0880$. One cannot do the same for the small range $0 < z < z_{*}$ because 
$s(z)$ changes sign from positive to negative in this region, whereas 
$Z = -\alpha^2 s^2$ is negative for all $z$. Hence $f_{y}(Z)$ takes values 
already used to fit redshifts for $z > z_{*}$. This means the very late time 
expansion history deviates from the $\Lambda$CDM model.

We can use the MOND cosmological equations (\ref{FLRWphi}), (\ref{FLRWxipsi})
and (\ref{Fried1}) to derive a system of local, first order differential 
equations for $\dot{\phi}$ and $H$ which can be evolved inward from $z = z_{*}$,
starting with the initial conditions $\dot{\phi}_{*} = 0$ and $H_{*} = H_0 
\widetilde{H}(z_*)$. Our labor is reduced by some preliminary rescalings,
\begin{equation}
\dot{\phi} \equiv 6 H_0 \!\times\! \Phi(z) \qquad , \qquad H \equiv H_0
\!\times\! h(z) \; .
\end{equation}
Then equation (\ref{FLRWphi}) implies,
\begin{equation}
\Phi' = \frac{3 \Phi}{1 \!+\! z} + h' - \frac{h}{1 \!+\! z} \; . 
\label{Phieqn}
\end{equation}
Dividing equation (\ref{Fried1}) by $3 H_0^2$ gives,
\begin{eqnarray}
\lefteqn{h^2(z) + \frac{6 f_y(Z)}{\alpha^2} + 12 h(z) \Phi(z) f'_{y}(Z) + 
24 h^2(z) \!\! \int_{z}^{\infty} \!\! dz' \, \frac{ \Phi(z') f'_{y}(Z')}{
(1 \!+\! z') h(z')} } \nonumber \\
& & \hspace{6cm} = \Omega_{\rm r} (1 \!+\! z)^4 + \Omega_{\rm b} (1 \!+\! z)^3 
+ \Omega_{\Lambda} \; . \qquad \label{Fried1a}
\end{eqnarray}
Dividing by $h^2(z)$ and differentiating with respect to $z$ will eliminate
the integration. Before giving the result it is worthwhile making some
simplifications based on the fact that the redshifts we seek to understand 
are very near $z = 0$. First, the radiation term to the right of 
(\ref{Fried1a}) can be dropped. Second, there is no point to using more than 
the first two terms of the small $s$ expansion (\ref{smalls}) for $f_{y}(Z)$,
\begin{eqnarray}
-\frac{f_{y}(Z)}{\alpha^2 \Omega_{\rm c}} & \!\!\!\! \simeq \!\!\!\! & 
\frac{(1 \!+\! z_{\Lambda})^3}{12} \Biggl\{ \frac{A \sqrt{-Z}}{\alpha 
\sqrt{\Omega_{\Lambda}}} - \frac{B Z}{\alpha^2 \Omega_{\Lambda}} 
\Biggr\} = \frac{(1 \!+\! z_{\Lambda})^3}{12} \Biggl\{ \tilde{a} \Phi + 
\tilde{b} \Phi^2\Biggr\} , \qquad \\
-\frac{f'_{y}(Z)}{\alpha^2 \Omega_{\rm c}} & \!\!\!\! \simeq \!\!\!\! & 
\frac{(1 \!+\! z_{\Lambda})^3}{12} \Biggl\{ -\frac{A}{2 \alpha 
\sqrt{-\Omega_{\Lambda} Z}} - \frac{B}{\alpha^2 \Omega_{\Lambda}} \Biggr\} 
= \frac{(1 \!+\! z_{\Lambda})^3}{12 \alpha^2} \Biggl\{ -\frac{\tilde{a}}{2 \Phi} 
- \tilde{b} \Biggr\} , \qquad
\end{eqnarray}
where we define $\tilde{a} \equiv \frac{A}{\sqrt{\Omega_{\Lambda}}} \simeq
-0.9192$ and $\tilde{b} \equiv \frac{B}{\Omega_{\Lambda}} \simeq +0.1843$.
It is also convenient to define $\tilde{c} \equiv \Omega_{\rm c} (1 + 
z_{\Lambda})^3 \simeq +0.5824$. With these simplifications the final evolution 
equation for $h(z)$ is,
\begin{eqnarray}
\lefteqn{ \Biggl\{ -\frac{\tilde{a}}{2 h} - \frac{\tilde{b} \Phi}{h} + 
\frac{\tilde{a} \Phi}{h^2} + \frac{2 [\Omega_{\rm b} (1 \!+\! z)^3 \!+\!
\Omega_{\Lambda}]}{ \tilde{c} h^2} \Biggr\} h' + \Biggl\{ -\frac{\tilde{a}}{2 h} 
+ \tilde{b} \Biggr\} \Phi' } \nonumber \\
& & \hspace{7cm} = \frac{\tilde{a} \!+\! 2 \tilde{b} \Phi}{1 \!+\! z}
+ \frac{3 \Omega_{\rm b} (1 \!+\! z)^2}{\tilde{c} h} \; . \qquad \label{Fried1b}
\end{eqnarray}

Figure~\ref{modelresults} shows the results of evolving equations 
(\ref{Phieqn}) and (\ref{Fried1b}) inward from $z = z_{*}$, starting with
the initial conditions $\Phi(z_{*}) = 0$ and $h(z_{*}) = \widetilde{H}(z_{*})
\simeq 1.043$. The evolution of $\Phi(z)$ is not significantly different from 
the $\Lambda$CDM model, however, the evolution of $h(z)$ differs markedly. 
Instead of continuing to decline to the $\Lambda$CDM value of $\widetilde{H}(0) 
= 1$, $h(z)$ turns around and increases slightly to $h(0) \simeq 1.045$. The 
increase is not large but halting the decrease may provide an explanation for 
the increasingly significant tension between inferences of $H_0$ based on 
data from large $z$ \cite{Ade:2015xua} (cosmic ray background anisotropies 
and baryon acoustic oscillations) and from small $z$ \cite{Riess:2016jrr} 
(Hubble plots),
\begin{eqnarray}
{\rm Large} \; z & \Longrightarrow & H_0 = \Bigl( 67.74 \pm 0.46
\Bigr)~\frac{\rm km}{\rm s~Mpc} \; , \label{largezH} \\
{\rm Small} \; z & \Longrightarrow & H_0 = \Bigl( 73.24 \pm 1.74
\Bigr)~\frac{\rm km}{\rm s~Mpc} \; . \label{smallzH}
\end{eqnarray}
The model was defined using the large $z$ numbers \cite{Ade:2015xua}, which
moves our prediction for the current Hubble parameter closer to the small 
$z$ result (\ref{smallzH}),
\begin{equation}
1.045 \times \Bigl( 67.74 \pm 0.46 \Bigr)~\frac{\rm km}{\rm s~Mpc}
= \Bigl( 70.79 \pm 0.48 \Bigr) \frac{\rm km}{\rm s~Mpc} \; .
\end{equation}

\section{Discussion}

This paper concerns a metric-based realization of MOND \cite{Deffayet:2011sk,
Deffayet:2014lba} whose Lagrangian (\ref{LMOND}) involves an algebraic function 
$f_{y}(Z)$ of a nonlocal scalar (\ref{Zdef}). For gravitationally bound systems 
$Z[g]$ is typically positive and the form of $f_{y}(Z)$ is well constrained by 
the Tully-Fisher relation, weak lensing and solar system tests. For cosmological 
systems the metric's temporal variation is typically more important than its 
spatial dependence, which causes $Z[g]$ to be negative. Our goal in this paper 
has been to determine how the function $f_{y}(Z)$ must depend upon negative $Z$ 
so as to reproduce the $\Lambda$CDM expansion history
without dark matter.

We did not quite succeed because specializing to the $\Lambda$CDM geometry
does not result in a one-to-one function $Z(z)$ for all $z > 0$. For large
$z$ the function $Z(z)$ is negative and its magnitude decreases as $z$ 
decreases. However, $Z(z)$ touches zero at $z_{*} \simeq 0.0880$ and then 
returns to negative values in the range $0 < z < z_{*}$. Therefore, we can
only choose $f_y(Z)$ to enforce the $\Lambda$CDM expansion history for
$z > z_{*}$. In this region we determined $f_y(Z)$ numerically, and then 
showed that a simple combination of analytic functions 
(\ref{goodlarge}-\ref{delta}) provides an excellent fit, with $Z \equiv -
\alpha^2 s^2$ and $f_y(Z) \equiv -\alpha^2 \Omega_{\rm c} f(z)$. We also 
demonstrated that the model's deviation from $\Lambda$CDM in the range 
$0 < z < z_{*}$ causes the current Hubble parameter to be about 4.5\% 
larger than for the $\Lambda$CDM model. This would reduce (from $3.2 
\sigma$ to only $1.4 \sigma$) the tension which currently exists between 
inferences of $H_0$ which are based on data from large $z$ \cite{Ade:2015xua} 
and those based on small $z$ data \cite{Riess:2016jrr}.

Cosmology offers an interesting venue for comparing nonlocal MOND with 
the $\Lambda$CDM model. Both models incorporate the known densities of 
radiation and baryonic matter, and both models assume an absurdly small 
cosmological constant. In both cases the remaining component of the 
Friedmann equation is abstracted to cosmology from an explanation for 
very large and weakly gravitationally bound structures. In both cases 
there is no definitive derivation of this remaining component from
fundamental theory although possibilities exist. One major difference
is that the $\Lambda$CDM model requires only the single free parameter
$\Omega_{\rm c}$ to describe the remaining component of the Friedmann 
equation whereas nonlocal cosmology has a free function $f_{y}(Z)$ 
for $Z < 0$. On the other hand, the function required does not seem
outlandish.

The most reasonable conclusion is probably that the cosmology of nonlocal 
MOND will stand or fall depending on what it predicts now that the 
function $f_y(Z)$ has been fixed. The model can be used to study things 
such as the response to recent disturbances of gravitationally bound 
systems and to perturbations around the background cosmology. The analogy 
is how numerically determining the distortion function \cite{Deffayet:2009ca} 
of a nonlocal model of dark energy \cite{Deser:2007jk} has facilitated 
detailed studies of structure formation in that model 
\cite{Park:2012cp,Dodelson:2013sma,Park:2016jym}. We do not yet know how 
these studies will turn out for the nonlocal realization of 
MOND,\footnote{Although one obvious point is that the model does reproduce 
the usual $\Lambda$CDM expansion history for all $z > z_{*}$, and the 
density of baryons is unchanged, so Big Bang Nucleosynthesis is not 
affected.} however, a crucially important point is that the function 
$f_y(Z)$ is NOT small for cosmology. This means that the MOND corrections 
have a reasonable chance to reproduce what dark matter does in general 
relativity.

The surprising (and wonderful) fact that MOND corrections are large for 
cosmology was not expected by many who attempted to guess the form a 
relativistic extension of MOND might take. These people made the reasonable 
assumption that MOND corrections to cosmology should be negligible because 
they are small for gravitationally bound systems, and because even those 
small corrections entirely disappear when the curvature reaches values far 
smaller than it has taken in cosmological history. This assumption fails 
for two reasons:
\begin{itemize}
\item{The invariant (\ref{Zdef}) becomes large in cosmology; and}
\item{The function $f_y(Z)$ is not suppressed for large negative $Z$ 
the way it is for large positive $Z$.}
\end{itemize}
The first point is a consequence of invariance. The phenomenology of MOND 
dictates (\ref{Zdef}) as the simplest form for $Z[g]$ \cite{Deffayet:2011sk,
Deffayet:2014lba}, and just evaluating this functional for the $\Lambda$CDM 
cosmology happens to produce numerically large results. Of course the 
second point resulted from how we choose to extend the function $f_{y}(Z)$
for negative $Z$. However, it is important to realize that this decision 
makes sense if one conceives of MOND as derived from the vacuum polarization 
of inflationary gravitons \cite{Woodard:2014wia}. Those gravitons have 
cosmological scales so it is entirely reasonable to expect strong effects 
on cosmological scales but only weak effects on smaller scales.

\centerline{\bf Acknowledgements}

We are grateful for conversations and correspondence with C. Deffayet,
S. Deser and G. Esposito-Farese. This work was partially supported by NSF 
grant PHY-1506513 and by the Institute for Fundamental Theory at the 
University of Florida.

\end{document}